\documentclass[a4paper,11pt]{article}
\pdfoutput=1 
\usepackage{jheppub}
\usepackage{graphicx}
\graphicspath{{./}{Figs/}}

\usepackage{amsfonts}
\usepackage{mathtools}
\usepackage{xcolor}
\usepackage{verbatim}

\newcommand{\benum}{\begin{enumerate}}
\newcommand{\eenum}{\end{enumerate}}
\newcommand{\bi}{\begin{itemize}}
\newcommand{\ei}{\end{itemize}}

\newcommand{\neff}{n}
\newcommand{\ceff}{C_{\rm eff}}

\newcommand{\be}{\begin{equation}}
\newcommand{\ee}{\end{equation}}
\newcommand{\beq}{\begin{eqnarray}}
\newcommand{\eeq}{\end{eqnarray}}
\newcommand{\bea}{\begin{eqnarray}}
\newcommand{\eea}{\end{eqnarray}}

\newcommand{\nnmb}{\nonumber}

\newcommand{\vbar}{\bar{v}}
\newcommand{\lrf}[2]{\left(\frac{#1}{#2}\right)}
\newcommand{\lrfsq}[2]{\left[\frac{#1}{#2}\right]}
\newcommand{\ttil}{\tilde{t}}
\newcommand{\atil}{\tilde{a}}

\newcommand{\Rmnum}[1]{\expandafter\@slowromancap\romannumeral #1@}

\def\kev{\,{\rm keV}}
\def\mev{\,{\rm MeV}}
\def\gev{\,{\rm GeV}}

\begin{document} 
  \begin{flushright}
    KCL-PH-TH/2018-47
%    ADP--17--08/T1014
  \end{flushright}

%%%%%%%%%%%%%%%%%%%%%%%%%%%%%%%%%%%%%%%%%%%%
 \title{\boldmath %
Probing the pre-BBN universe with gravitational waves from cosmic strings}
 %%%%%%%%%%%%%  Authors  Alphabetically %%%%%%%%%%%%%%%%%%
\author[a]{ Yanou Cui,}%\note{ORCID ID: }}
\author[b,c]{Marek  Lewicki,}%\note{ORCID ID: 0000-0002-8378-0107}}
\author[d]{David E.  Morrissey,}%\note{ORCID ID: 0000-0002-8378-0107}}
\author[e,f]{James D. Wells,}%\note{ORCID ID: 0000-0002-8943-5718}}
%%%%%%%%%%%%%  Affiliations %%%%%%%%%%%%%%%%%%%%%%%%%
\affiliation[a]{Department of Physics and Astronomy, University of California,
900 University Avenue, Riverside, CA 92521, USA}
\affiliation[b]{Kings College London, Strand, London, WC2R 2LS, United Kingdom}
%\affiliation[b]{ARC Centre  of  Excellence  for Particle  Physics  at  the  Terascale
% (CoEPP) \& CSSM, Department of Physics, University of Adelaide, South Australia 5005, Australia}
\affiliation[c]{Faculty of Physics, University of Warsaw ul.\ Pasteura 5, 02-093 Warsaw, Poland}
\affiliation[d]{TRIUMF, 4004 Wesbrook Mall, Vancouver, BC, Canada V6T 2A3}
\affiliation[e]{Leinweber Center for Theoretical Physics, University of Michigan,
450 Church Street, Ann Arbor, MI 48109, USA} 
\affiliation[f]{Theory Group, Deutsches Elektronen-Synchrotron DESY, Theory Group, D-22603 Hamburg, Germany}
%%%%%%%%%%%%%%%%  Email Addresses %%%%%%%%%%%%%%%%%%%%
\emailAdd{yanou.cui@ucr.edu}
\emailAdd{marek.lewicki@kcl.ac.uk}
\emailAdd{dmorri@triumf.ca}
\emailAdd{jwells@umich.edu}
%%%%%%%%%%%%%%%%%%%%%%%%%%%%%%%%%%%%%%%%%%%%
%  Abstract
\abstract{
Many motivated extensions of the Standard Model predict the existence of cosmic strings.  Gravitational waves originating from the dynamics of the resulting cosmic string network have the ability to probe many otherwise inaccessible properties of the early universe. In this study we show how the spectrum of gravitational waves from a cosmic string network can be used to test the equation of state of the early universe prior to Big Bang Nucleosynthesis~(BBN).  We also demonstrate that current and planned gravitational wave detectors such as LIGO, LISA, DECIGO/BBO, and ET/CE have the potential to detect signals of a non-standard pre-BBN equation of state and evolution of the early universe (e.g., early non-standard matter domination or kination domination) or new degrees of freedom active in the early universe beyond the sensitivity of terrestrial collider experiments and cosmic microwave background measurements.
} 
%\keywords{cosmic string, gravitational waves, modified cosmology}

\maketitle
\flushbottom

%%%%%%%%%%%%%%%%%%%%%%%%%%%%%%%%%%%%%%%%%%%%%%%%%%%%%%%%%%%%%%%%%%%%%%%

\section{Introduction} \label{sec:intro}

Remarkable progress has been made in understanding the universe through
detailed observations of the electromagnetic radiation 
emitted by the cosmos.  These measurements, spanning a range of frequencies 
from radio to gamma-ray~\cite{Hill:2018trh}, 
have led to the $\Lambda$CDM model of cosmology
in which the universe is currently dominated by dark energy and cold dark
matter with smaller components of baryonic matter 
and radiation~\cite{Aghanim:2018eyx}.

  Extrapolating the $\Lambda$CDM model back in time suggests that the
very early universe was dominated by radiation in the form of photons
and other relativistic particles.  This extrapolation is strongly
supported by measurements of the cosmic microwave background~(CMB),
corresponding to the photons that escaped after recombination
when the radiation temperature was about $0.3\;\text{eV}$~\cite{Aghanim:2018eyx}.
The success of Big Bang Nucleosynthesis~(BBN) in predicting primordial light
element abundances gives additional convincing evidence for early radiation 
domination~(RD) up to temperatures close to 
$T \simeq 5\,\mev$~\cite{Kawasaki:1999na,Kawasaki:2000en,Hannestad:2004px}.  
Going even further back, the observed flatness and uniformity of 
the cosmos and the spectrum of density perturbations
suggest that this radiation era was preceded by a period of rapid
expansion such as inflation~\cite{Guth:1980zm,Linde:1981mu,Albrecht:1982wi,Akrami:2018odb}.

Very little is known empirically about the state of the universe between
the end of inflation and the start of BBN~\cite{Boyle:2007zx}.
A minimal assumption is that inflation was followed by reheating  
to a very hot radiation phase with temperature $T \gg {\rm TeV}$, 
which then cooled adiabatically until giving way to the recent matter 
and dark energy phases.  We refer to this scenario, with only radiation 
domination~(RD) over many orders of magnitude in temperature between 
reheating and the matter epoch, as the standard cosmology~\cite{Kolb:1990vq}.  
While this assumption is made frequently (and often implicitly),
it has not been tested directly.  Furthermore, non-standard cosmological
scenarios with an extended period of domination by something other
than radiation between inflation and BBN have strong motivation 
from many perspectives, including dark matter, axions, string compactification, reheating, and baryogenesis~\cite{Moroi:1999zb, Salati:2002md, Boyle:2007zx, Chung:2007vz, Gelmini:2008sh, Visinelli:2009kt,Erickcek:2015jza, Dutta:2016htz,Giblin:2017wlo, Visinelli:2017qga, Visinelli:2017imh, Visinelli:2018wza, Poulin:2018dzj, Dutta:2018zkg, Redmond:2018xty, Nelson:2018via}.
Testing the paradigm of pre-BBN cosmology is therefore of great significance 
in advancing our understanding of the universe.

Gravitational waves~(GWs) may provide a means of looking back in time beyond 
the BBN epoch and probing the universe in its very early stages~\cite{Allen:1996vm, Boyle:2007zx,Cui:2017ufi,Caprini:2018mtu}.  
The observation of binary mergers by the LIGO/Virgo collaboration has 
already given further support to the $\Lambda$CDM cosmology~\cite{Abbott:2017xzu}, although -- and this is important to the motivation of this work -- the GWs observed were created only relatively recently.  Opportunity to look even 
further back in time with GWs arises because, in contrast to photons, 
GWs free-stream throughout the entire history of the cosmos. 
Indeed, GWs emitted as far back as inflation could potentially 
be detected by LIGO/Virgo~\cite{Aasi:2014mqd} 
or proposed future detectors such as LISA~\cite{Audley:2017drz}, 
BBO/DECIGO~\cite{Yagi:2011wg}, the 
Einstein Telescope~(ET)~\cite{Punturo:2010zz,Hild:2010id},
and Cosmic Explorer~(CE)~\cite{Evans:2016mbw}.

A stable and predictable source of primordial GWs is needed
if they are to be used to test very early cosmology.
Two promising and well-motivated potential sources are cosmic strings~\cite{Vilenkin:1984ib,Caldwell:1991jj,Hindmarsh:1994re,Vilenkin:2000jqa}
and primordial inflation~\cite{Grishchuk:1974ny,Starobinsky:1979ty,Allen:1987bk}.
The application of inflationary GWs to probe the expansion history
of the universe was studied in Refs.~\cite{Turner:1993vb,Seto:2003kc,Nakayama:2008wy,Nakayama:2008ip} and for non-standard histories 
in Refs.~\cite{Giovannini:1998bp,Riazuelo:2000fc,Sahni:2001qp,Tashiro:2003qp}.
However, current limits from CMB isotropy typically push the inflationary
stochastic GW spectra below the sensitivity of current and 
next generation detectors~\cite{Guzzetti:2016mkm,Ananda:2006af,Smith:2005mm} 
(although see Ref.~\cite{Lasky:2015lej}).  Thus, we focus on GWs from 
cosmic strings in this work.

Cosmic strings are approximately one-dimensional objects of macroscopic length 
that arise nearly generically in theories of physics beyond the 
Standard Model~(SM).  Specific examples include topologically-stable 
field configurations in theories with a spontaneously broken 
$U(1)$ symmetry~\cite{Nielsen:1973cs,Kibble:1976sj}, 
as well as fundamental or composite strings in superstring 
theory~\cite{Copeland:2003bj,Dvali:2003zj, Polchinski:2004ia,Jackson:2004zg,Tye:2005fn}. 
Their macroscopic properties are mostly
characterized by their tension (energy per unit length) $\mu$, which is
typically on the order of the square of the energy scale of new physics 
that gives rise to them, and directly constrained by CMB measurements 
to $G\mu < 1.1\times 10^{-7}$~\cite{Charnock:2016nzm},
where $G$ is Newton's constant.

  Cosmic strings emit gravitational radiation as part of their cosmological
evolution~\cite{Vilenkin:1981bx,Vachaspati:1984gt,Turok:1984cn,Burden:1985md}.  
After formation, cosmic strings are expected to quickly reach 
a scaling regime in which their net energy density tracks the total 
cosmological energy density with a relative fraction $G\mu$~\cite{Albrecht:1984xv,Bennett:1987vf,Allen:1990tv}.  
This regime consists of a small number of Hubble-length long strings 
and a collection of many closed string loops.  As the universe expands, 
the long strings intersect and intercommute to form new loops, 
while the existing loops oscillate and emit radiation, including GWs.
This continual transfer of energy from long strings to loops to  
radiation is essential for the string network density to track the
total energy density of the universe, rather than becoming dominant
like other topological defects such as 
monopoles~\cite{Zeldovich:1978wj,Preskill:1979zi} and 
domain walls~\cite{Zeldovich:1974uw}.
In particular, the presence of cosmic strings with small $G\mu \ll 1$ 
need not disrupt the standard cosmology.

  For many classes of cosmic strings, the dominant radiation emission
is in the form of GWs.  This is true for ideal Nambu-Goto strings,
many types of cosmic strings emerging from superstring theory,
and possibly those created by local $U(1)$ symmetry 
breaking~\cite{Olum:1999sg,Moore:2001px} 
(although see Refs.~\cite{Vincent:1997cx,Bevis:2006mj,Figueroa:2012kw,Helfer:2018qgv}
for arguments that local strings emit mainly massive vector and Higgs
quanta instead).  
In contrast, cosmic strings derived from global symmetry breaking 
are expected to radiate mainly to light Goldstone 
quanta~\cite{Srednicki:1986xg,Vilenkin:1986ku,Damour:1996pv,Cui:2008bd,Long:2014mxa}, with much weaker emission to GWs.  
We focus on cosmic strings that radiate significantly 
to GWs through loop formation and emission in this work.

  The GW frequency spectrum from a cosmic string network 
is sensitive to the evolution of the cosmos when the GWs were emitted.
In any given frequency band observed today, the dominant contribution
to the signal comes from loops emitting GWs at a specific time in the early
universe~\cite{Caldwell:1991jj,Allen:1996vm,Cui:2017ufi}.  As a result of this 
frequency-time relation, we show that the cosmological equation 
of state leaves a distinct imprint on the frequency spectrum 
of GWs from cosmic strings.  Moreover, the portion of the spectrum
from loops formed and emitting during RD has a distinctive nearly flat plateau
with a substantial amplitude over many decades in frequency.
Measuring the GW signal from a cosmic string network over a range 
of frequencies could therefore provide a unique picture of the very 
early universe that could potentially expand back before the era of BBN.

  The outline of this paper is as follows.  After this introduction,
we review cosmic string scaling and derive the GW frequency spectrum
from a string network in Section~\ref{sec:spectrum}.  We also exhibit
the relationship between the GW spectrum and the loop emission rates 
and formation times, and apply these to the concurrent background cosmology.  
In Section~\ref{sec:cosmomap} we show how this relationship together 
with the anticipated sensitivities of current and planned GW detectors 
can be used to test the standard cosmological scenario as well as deviations 
from it, including large numbers of additional (massive) degrees of freedom 
and modified equations of state.  
Some of the challenges to detecting these GW signals and identifying them as 
coming from cosmic strings, and ways to overcome them, 
are discussed in Sec.~\ref{sec:detect}.
Finally, Section~\ref{sec:conc} is reserved for our conclusions.

  The results in this paper expand upon those of our previous study in
Ref.~\cite{Cui:2017ufi}.  Relative to the work, we present in great detail
the time-frequency connection of cosmic string GWs and its relation 
to the background cosmology.  We also expand significantly on the experimental 
sensitivity of GW probes to new degrees of freedom active during 
early universe with presence of cosmic string dynamics, and extend our
study of standard and modified cosmological histories.

\section{GW spectrum of a cosmic string network}
\label{sec:spectrum}

In this section we derive the GW frequency spectrum from a cosmic string network. 
We assume a network of ideal Nambu-Goto cosmic strings with unit
reconnection probability and dominant energy loss through loop formation and
emission of gravitational radiation.  

\subsection{Cosmic string scaling and loop production}

  Cosmic string scaling is achieved through a balance of the 
slow $a^{-2}$ dilution of the horizon-length long-string density 
and the transfer of energy out of the long-string network by the 
production of closed string loops~\cite{Vilenkin:2000jqa}.  
These loops oscillate, emit energy in gravitational radiation, 
and eventually decay away.  To compute the GW spectrum from these processes, 
estimates  are needed for the sizes and rates of the loops formed 
by the long string network.
  
  Recent cosmic string simulations find that a fraction of about 10\%  
of the energy transferred by the long strings to loops is in the form 
of relatively large loops, with the remaining 90\% going to highly boosted 
small loops~\cite{Vanchurin:2005pa,Martins:2005es,Olum:2006ix,Ringeval:2005kr,BlancoPillado:2011dq,Blanco-Pillado:2017oxo}.  
The large loops give the dominant contribution
to the GW signal and we focus exclusively on them, since the relativistic 
small loops lose most of their energy to simple redshifting.  
Large loops are formed with a characteristic initial 
length $l_i$ that tracks the time $t_i$ of formation,
\beq
l_i = \alpha t_i \ ,
\eeq
where $\alpha$ is an approximately constant loop size 
parameter~\cite{Vanchurin:2005pa,Olum:2006ix,Martins:2005es,Ringeval:2005kr,BlancoPillado:2011dq,Blanco-Pillado:2013qja,Blanco-Pillado:2017oxo}. 
We make the simplifying assumption of monochromatic (large) loop formation
with $\alpha = 0.1$, which gives a good reproduction of the loop size
distribution of Refs.~\cite{Blanco-Pillado:2013qja,Blanco-Pillado:2017oxo}.
We comment on the impact of modifying the value of $\alpha$ in Section~\ref{sec:alphasensitivity}. 

  The formation rate of (large) loops by a scaling string network
is also needed to compute the GW spectrum.  For this, we
use the velocity-dependent one-scale~(VOS) model to describe
the properties of the long string network in the scaling 
regime~\cite{Martins:1995tg, Martins:1996jp,Martins:2000cs,Avelino:2012qy,Sousa:2013aaa},
and we match the rate of energy release by the long string network needed
to maintain scaling with the rate of energy going to 
loops~\cite{Vilenkin:2000jqa}.
The VOS model describes the long string network in terms of a characteristic
length (as a fraction of the horizon) $\xi$ and a mean string velocity $\bar{v}$,
and is found to give a good analytic description of the network properties
during scaling.  Let us emphasize, however, that we only use the VOS model 
to describe the long string network; we base the structure of the loops 
on the results of direct simulations~\cite{Blanco-Pillado:2013qja,Blanco-Pillado:2017oxo}.

  Consider a scaling network evolving in a cosmological background
driven by a dominant source of energy density that dilutes according to
\beq
\rho ~\propto~ a^{-n} \ .
\label{eq:neff}
\eeq
This implies $a(t) ~\propto~ t^{2/n}$,
with $n=3,\,4$ giving the familiar cases of matter and radiation domination.  
Within such a background, the VOS model describes the long string
network in terms of a characteristic string velocity $\vbar$ and length
parameter $\xi$ given by~\cite{Martins:1995tg, Martins:1996jp,Martins:2000cs}
\begin{equation}\label{eq:loopgamma}
\xi =
\frac{\neff}{2}\sqrt{\frac{ k(\bar{v})[k(\bar{v})+\bar{c}]}{2\left(\neff-2\right)}} \ , \quad
\bar{v}=
\sqrt{\frac{\neff}{2}\frac{k(\bar{v})}{[k(\bar{v})+\bar{c}]}\left(1-\frac{2}{\neff}\right)} , 
\end{equation} 
where $\bar{c}$ is a loop chopping efficiency parameter and $k(\vbar)$
is a function of $\vbar$ to be determined.  We fix  
$\bar{c} = 0.23$ based on numerical simulations~\cite{Martins:2000cs},
and we use the ansatz of Ref.~\cite{Martins:2000cs} for the function $k(\vbar)$:
\begin{equation}\label{eq:loopv}
k(\bar{v})=
\frac{2\sqrt{2}}{\pi}
\left( 1-\bar{v}^2\right)
\left( 1+2\sqrt{2}\bar{v}^3\right)
\frac{1-8\bar{v}^6}{1+8\bar{v}^6} \ .
\end{equation}
In terms of $\xi$ and $\vbar$, the energy density $\rho_L$
of the long string network is
\beq
\rho_{L} = \frac{\mu}{(\xi\,t)^2} \ ,
\eeq
while the rate of energy loss needed to maintain scaling is
\beq
\frac{d\rho_{L}}{dt} = \bar{c}\,\vbar\,\frac{\rho_{L}}{(\xi\,t)} \ .
\label{eq:longloss}
\eeq

  To estimate the loop formation rate, we identify the energy 
loss rate of Eq.~\eqref{eq:longloss} with the rate of energy transferred
to loops.  If large loops of initial size $l_i = \alpha\,t_i$
and Lorentz boost $\gamma$ make up a fraction $\mathcal{F}_{\alpha}$
of the energy released by long strings, their formation rate is 
\beq
\frac{dn_{\alpha}}{dt} 
= \mathcal{F}_{\alpha}\frac{C_{eff}}{\alpha}\,t^{-4} \ ,
\label{eq:looprate}
\eeq
with
\beq
C_{eff} = \frac{\tilde{c}}{\gamma}\,\vbar\,\xi^{-3} \ . 
\eeq
Recent simulations find $\alpha \simeq 0.1$, $\mathcal{F}_{\alpha} \simeq 0.1$, 
and $\gamma \simeq \sqrt{2}$, and we use these as default values
in the analysis to follow~\cite{Blanco-Pillado:2013qja,Blanco-Pillado:2017oxo}.

\begin{figure}
\centering
\includegraphics[height=6cm]{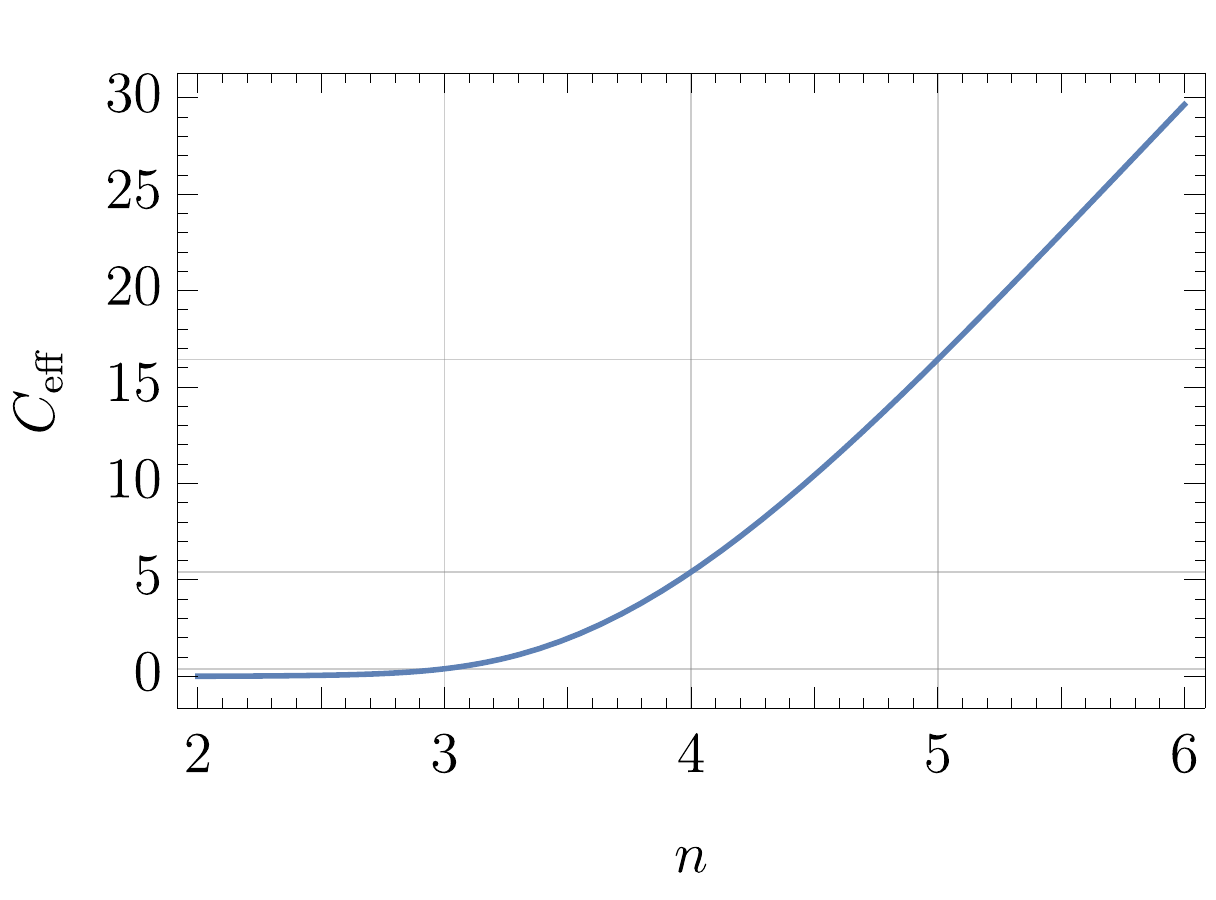}
\caption{
Dependence of the loop emission factor $\ceff$ derived from the VOS model
on the cosmological energy density redshift parameter $n$ defined 
in Eq.~\eqref{eq:neff}.
\label{fig:Cofn}}
\end{figure}

  In Fig.~\ref{fig:Cofn} we show the result for $C_{eff}$ as a function
of the background cosmology scaling factor over the range $n\in[2, 6]$.  
We find $C_{eff} = 0.39$ and $5.4$ during matter~($n=3$) and radiation~($n=4$) 
domination, respectively, which compare well with $C_{eff} = 0.5$ and $5.7$ 
found in detailed lattice simulations~\cite{BlancoPillado:2011dq,Blanco-Pillado:2013qja,Blanco-Pillado:2017oxo}.  The method used here can be applied to other 
cosmological backgrounds, and in particular we note that
$C_{eff}(n=6) \simeq 30.4$.

%%%%%%%%%%%%%%%%%%%%%%%%%%%%%%%%%%%%%%%%%%%%%%%%%%%%%%%%%%%%%%%%%%%%%%%%%%%%%%%%%%%%%%%%%%%%%
%%%%%%%%%%%%%%%%%%%%%%%%%%%%%%%%%%%%%%%%%%%%%%%%%%%%%%%%%%%%%%%%%%%%%%%%%%%%%%%%%%%%%%%%%%%%%%%

\subsection{Derivation of the GW frequency spectrum}

After formation, loops are found to emit energy in the form of gravitational
radiation at a constant rate
\be
\frac{d E}{d t}=-\Gamma G \mu^2 \ ,
\label{eq:eloss}
\ee  
where $\Gamma\approx 50$ is a dimensionless constant~\cite{
Vilenkin:1981bx,Turok:1984cn,Quashnock:1990wv,Blanco-Pillado:2013qja,Blanco-Pillado:2017oxo}. Thus, the length of a loop with initial size $l_i=\alpha t_i$
decreases as
\be 
\label{eq:loopsize}
l(t) =\alpha t_i -\Gamma G \mu \left(t-t_i \right) \ , 
\ee
until the loop disappears completely.
The total energy loss from a loop can be decomposed into a set of normal-mode 
oscillations at frequencies $\tilde{f}_{k}=2k/l$, 
where $k=1,2,3...$ is the mode number.
The relative emission rate per mode is found to scale with $k^{-4/3}$
and is given by~\cite{Blanco-Pillado:2013qja,Blanco-Pillado:2017oxo}
\begin{equation}\label{eq:Gammak}
\Gamma^{(k)} = \frac{\Gamma k^{-\frac{4}{3}}}{\sum_{m=1}^{\infty} m^{-\frac{4}{3}} } \ .
\end{equation}
Note that $\sum_{m=1}^{\infty} m^{-\frac{4}{3}} \simeq 3.60$ and
$\sum_k \Gamma^{(k)} = \Gamma$.  After emission at time $\tilde{t}$
the frequency of the GW redshifts, so the frequency observed today
is $f = [a(\tilde{t})/a(t_0)]\tilde{f}$.

  Combining the GW emission rate per loop of Eq.~\eqref{eq:eloss},
the emitted frequencies of Eq.~\eqref{eq:Gammak}, and the rate of 
(large $\alpha = 0.1$) loop formation of Eq.~\eqref{eq:looprate},
we can compute the relic GW background from a cosmic string network.
It is conventional to express this background in terms of
\beq
\Omega_{\rm GW} = \frac{f}{\rho_c}\frac{d\rho_{\rm GW}}{df} \ ,
\eeq
where $\rho_{\rm GW}$ is the energy density of GWs, $f$ is the frequency today,
and $\rho_c = 3H_0^2/8\pi G$ is the critical density.
Summing over all mode contributions,
\beq
\label{eq:GWdensity1}
\Omega_{\rm GW}(f) =\sum_k \Omega_{\rm GW}^{(k)}(f)  \ ,
\eeq
with 
\beq
\label{eq:GWdensity2}
\Omega_{\rm GW}^{(k)}(f) =
\frac{1}{\rho_c}
\frac{2k}{f}
\frac{\mathcal{F}_{\alpha}\,\Gamma^{(k)}G\mu^2}
{\alpha\left( \alpha+\Gamma G\mu\right)}
\int_{t_F}^{t_0}\!d\tilde{t}\;
\frac{\ceff(t_i^{(k)})}{t_i^{(k)\,4}} 
\bigg[\frac{a(\tilde{t})}{a(t_0)}\bigg]^5
\bigg[\frac{a(t^{(k)}_i)}{a(\tilde{t})}\bigg]^3
\,\Theta(t_i^{(k)} - t_F)~~~~~
\eeq
where the integral runs over the GW emission time $\tilde{t}$, and
\begin{equation}\label{eq:ti}
t_i^{(k)}(\tilde{t},f) = \frac{1}{\alpha+\Gamma G\mu}\left[ 
\frac{2 k}{f}\frac{a(\tilde{t})}{a(t_0)} + \Gamma G\mu\,\tilde{t}
\right].
\end{equation}
is the formation time of loops contributing with mode number $k$, 
and $t_F$ is the time at which the cosmic string network reached scaling, shortly after the formation of the network.
Note that the sum in Eq.~\eqref{eq:GWdensity1} is easily evaluated 
because 
$\Omega_{\rm GW}^{(k)}(f) 
= \frac{\Gamma^{(k)}}{\Gamma^{(1)}}\,\Omega_{\rm GW}^{(1)}(f/k)
=k^{-4/3}\,\Omega_{\rm GW}^{(1)}(f/k)$.
%$\Omega_{\rm GW}^{(k)}(f) =\frac{\Gamma^{(k)}}{\Gamma^{(1)}}\,\Omega^{(1)}(f/k)$.

%%%%%%%%%%%%%%%%%%%%%%%%%%%%%%%%%%%%%%%%%%%%%%%%%%%%%%%%%%%%%%%%%%%%%%
\begin{figure}
\centering
\includegraphics[height=8cm]{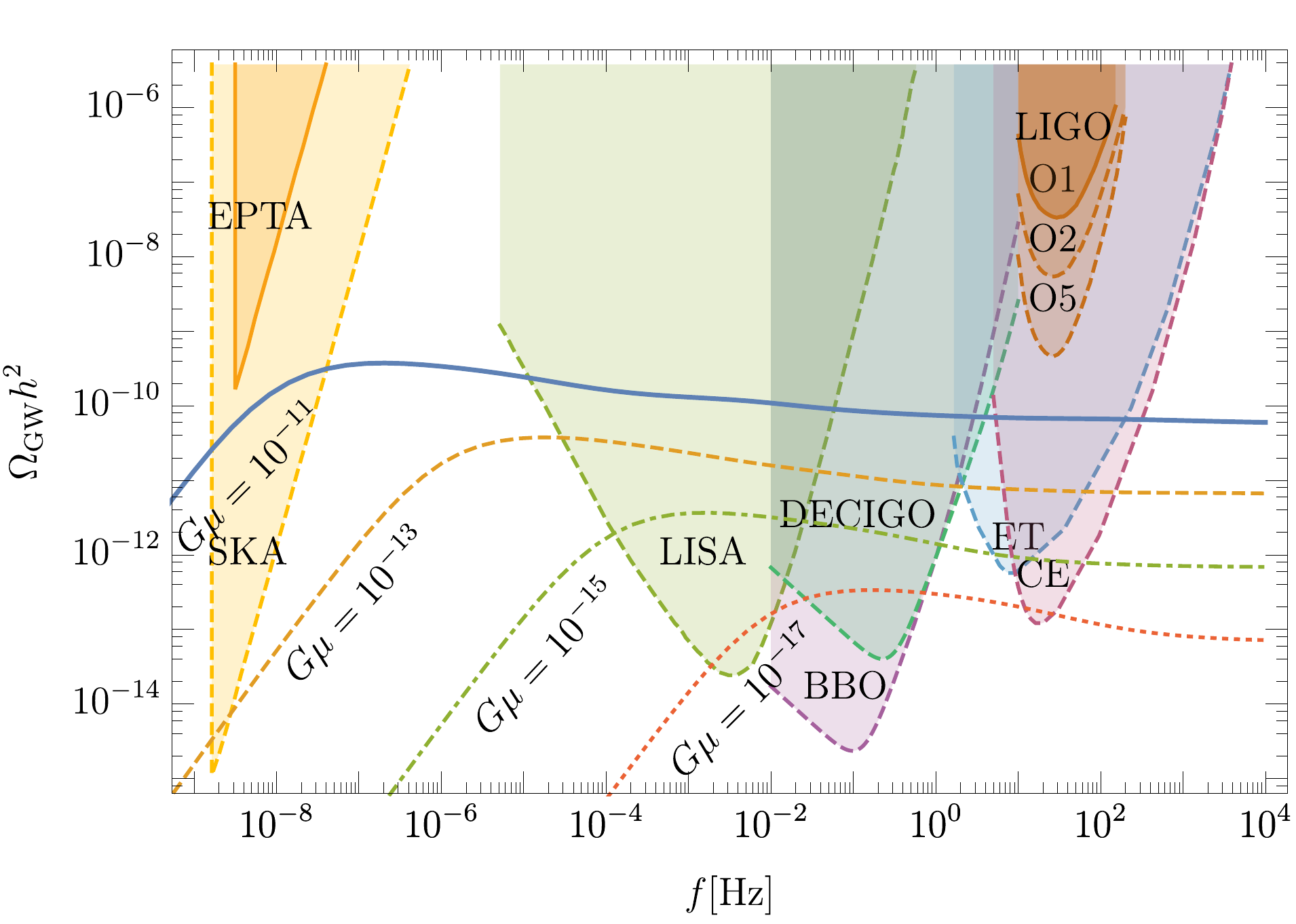}
\caption{
Gravitational wave spectrum from a cosmic string network with $\alpha = 0.1$
and $G\mu = 10^{-11},\,10^{-13},\,10^{-15},\,10^{-17}$.  Also shown are
the current sensitivities of LIGO and EPTA~(solid bounded regions), 
and the projected future sensitivities of LISA, DECIGO/BBO, ET/CE, 
and SKA~(dash bounded regions).
\label{fig:GWGmuplot}}
\end{figure}
%%%%%%%%%%%%%%%%%%%%%%%%%%%%%%%%%%%%%%%%%%%%%%%%%%%%%%%%%%%%%%%%%%%%%%%

   We show in Fig.~\ref{fig:GWGmuplot} the GW spectrum from a cosmic string
network with $\alpha = 0.1$, $G\mu = 10^{-11},\,10^{-13},\,10^{-15},\,10^{-17}$,
assuming standard cosmological evolution.
Also shown are the current and future sensitivities
of LIGO~\cite{TheLIGOScientific:2014jea,Thrane:2013oya,TheLIGOScientific:2016wyq,Abbott:2017mem}, 
and the projected sensitivities of LISA~\cite{Bartolo:2016ami}, 
DECIGO/BBO~\cite{Yagi:2011wg},
Einstein Telescope~(ET)~\cite{Punturo:2010zz,Hild:2010id},
and Cosmic Explorer~(CE)~\cite{Evans:2016mbw}.
The solid triangle in the upper left of the plot indicates the current limit 
from the European Pulsar Timing Array~(EPTA)~\cite{vanHaasteren:2011ni}, 
and the expected sensitivity of the future Square Kilometre 
Array~(SKA)~\cite{Janssen:2014dka}.  We see that the strongest current bound
on these GW spectra comes from EPTA and implies $G\mu \lesssim 2\times 10^{-11}$. 
Other recent estimates of the GW spectrum from a scaling cosmic string network
relative to current and future searches includes Refs.~\cite{Blanco-Pillado:2017oxo,Blanco-Pillado:2017rnf,Ringeval:2017eww}.

\subsection{Connecting GW frequencies to loop formation and emission times}
\label{sec:Tfrelation}

  The GW spectra shown in Fig.~\ref{fig:GWGmuplot} all share a characteristic
shape, with a dropoff at lower frequencies and a flattening at higher ones.
This shape is related to the cosmological background evolution when the loops 
contributing to a given frequency were formed and emitted 
GWs~\cite{Allen:1996vm}. 
In this section, we connect the GW frequency seen today to the time 
at which the dominant contribution to that frequency was emitted by 
the string network.  Later, we show how this connection can 
be used to test the evolution of the very early universe.

We begin with a simple analytic estimate of the frequency-time connection.
(See also Ref.~\cite{Kuroyanagi:2012jf}.)
For this, it is sufficient to focus exclusively on the $k=1$ mode 
which we find to be the dominant one in the cases of interest.  
We also set $t_F \to 0$ for now, and return to non-zero values later on.
The expression of Eq.~\eqref{eq:GWdensity2} involves an integral over
the GW emission time $\ttil$, with the contribution to the signal
over the time interval $(\ttil,\,\ttil+d\ttil)$ proportional to 
\beq
d\Omega_{\rm GW}(f) ~\propto~ d\ttil\times\mathcal{I}(\ttil,f) ~\equiv~
d\ttil\times \frac{1}{f}\,t_i^{-4}\bigg(\frac{a_i}{\atil}\bigg)^3\bigg(\frac{\atil}{a_0}\bigg)^5 \ ,
\eeq
with $a_i = a(t_i)$, $\atil = a(\ttil)$, and $t_i$ given by 
Eq.~\eqref{eq:ti} with $k=1$.  The $\ttil$ and $f$ dependence of 
the function $\mathcal{I}(\ttil,f)$ is approximately power-law,
and depends on the dominant cosmological background energy source
at times $\ttil$ and $t_i$.  Let us define the time $\ttil_M(f)$ to
be the value of $\ttil$ when the two terms in $t_i$ are equal in size:
\beq
\frac{2}{f}\frac{a(\ttil_M)}{a_0} = \Gamma G\mu\,\ttil_M \ . 
\label{eq:tm}
\eeq
If the background energy redshifts as $\rho(t_i) \propto a^{-m}$ at time $t_i$ 
and $\rho(\ttil) \propto a^{-n}$ at time $\ttil$, the approximate $\ttil$ 
and $f$ dependence of $\mathcal{I}(\ttil,f)$ is
\beq
\mathcal{I}(\ttil,f) ~\propto~
\left\{
\begin{array}{lcc}
f^{(3m-6)/m}\;\ttil^{(2/n)(6-4m)/m + 4/n}&~;~&\ttil < \ttil_M\\
&&\\
f^{-1}\;\ttil^{(6-4m)/m + 4/n}&~;~&\ttil \geq \ttil_M
\end{array}\right.
\label{eq:iscale}
\eeq
Integrating this power-law form is straightforward,
with the indefinite integral scaling according to
\beq
\int\!d\ttil\;\mathcal{I}(\ttil,f) ~\propto~ \ttil^{\,p}
\label{eq:powers1}
\eeq
with
\beq
p = \left\{
\begin{array}{lcl}
p_1 = 1+ (2/n)(6-4m)/m + 4/n&~~;~~&\ttil < \ttil_M\\
p_2 = 1 + (6-4m)/m +4/n&~~;~~& \ttil \geq \ttil_M
\end{array}
\right.
\label{eq:powers2}
\eeq
To evaluate Eq.~\eqref{eq:GWdensity2} in this approximation,
we divide the integral over $\ttil$ into non-overlapping regions
with distinct $(m,n)$ values and sum the piecewise contributions
of the form of Eqs.~\eqref{eq:powers1} and \eqref{eq:powers2}.
The power $p_1$ is positive for the ranges of interest $m,\,n \in (2,6]$,
implying that the contribution to the definite integral from 
$\ttil < \ttil_M$ is dominated by $\ttil \sim \min\{\ttil_M,t_0\}$.  
If $\ttil_M < t_0$ and $p_2 < 0$, which is true for most cases of interest 
in this work, the contribution from $\ttil > \ttil_M$ is also dominated 
by $\ttil \sim \ttil_M$ and has the same parametric size as that 
from $\ttil < \ttil_M$.  
In contrast, for $p_2 > 0$ the integral is dominated by the largest 
value of $\ttil$ in the corresponding $(m,n)$ region.

  The result of Eq.~\eqref{eq:iscale} can also be used to derive the
approximate frequency dependence of $\Omega_{\rm GW}(f)$.
We find
\beq
\Omega_{\rm GW}(f) &\propto& 
\left\{
\begin{array}{lcl}
f^{(3m-6)/m} &~;~& \ttil_M \geq t_0\\
%f^{\frac{2(mn - m - 3n)}{m(n-2)}} 
f^{2(mn-m-3n)/m(n-2)}
&~;~& \ttil_M < t_0,\; p_2 < 0\\
f^{-1} &~;~& \ttil_M < t_0,\; p_2 \geq 0
\end{array}
\right.\ ,
\label{eq:fscale}
\eeq
where $(m,n)$ refer to the cosmological scalings specifically
at $t_i(\ttil,f)$ and $\ttil$ for $\ttil = \min\{\ttil_M(f),\,t_0\}$.
For loop formation and GW emission in the radiation era, $(m,\,n) \simeq (4,\,4)$
and $\Omega_{\rm GW} \propto f^0$, corresponding to the flat plateaus seen
at higher frequencies in Fig.~\ref{fig:GWGmuplot}.  For loop formation
in the radiation era and GW emission in the matter era, $(m,\,n)= (4,\,3)$ 
giving $\Omega_{\rm GW} \propto f^{-1/2}$, which coincides with the decrease
seen in Fig.~\ref{fig:GWGmuplot} prior to the flat plateau.
The rising spectrum at low frequencies corresponds to $\ttil_M(f) > t_0$
(and $t_i < t_{eq}$ unless $f$ is very small), for which the dominant emission 
occurs around $\ttil \sim t_0$ implying $\Omega_{\rm GW}(f) \propto f^{3/2}$.  

\begin{figure}
\centering
\includegraphics[height=8cm]{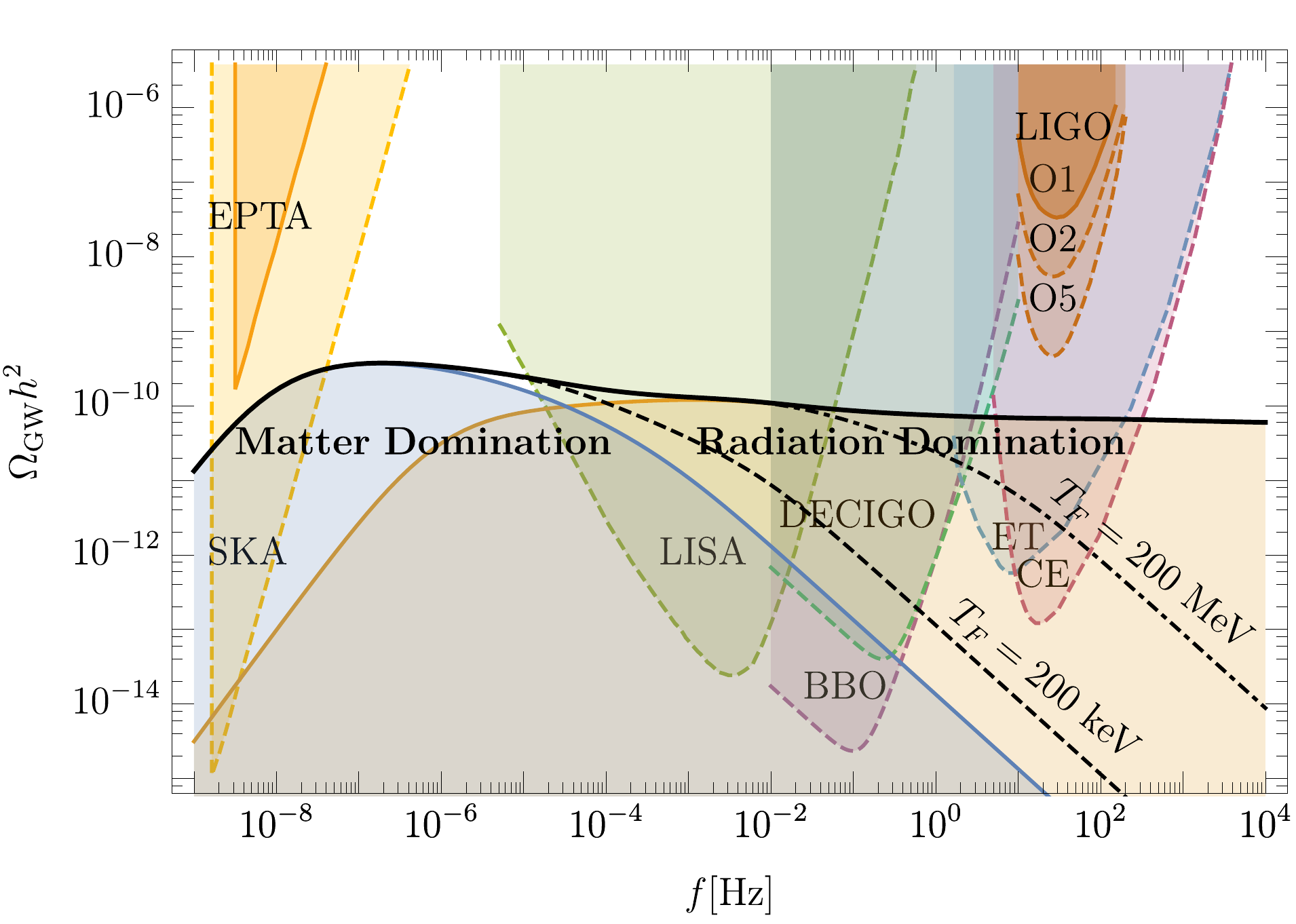}
\caption{Gravitational wave spectrum from a cosmic string network with 
$G\mu =10^{-11}$ and $\alpha = 10^{-1}$. 
The solid black line shows the full spectrum, while the blue shaded region 
shows the contribution from loops emitted during matter domination 
and the orange shaded region indicates the contribution from loops emitting
in the radiation era.  The black dashed lines show the effect on the spectrum
if only those loops formed at temperatures below $T_F = 200\,\kev,\,200\,\mev$
are included.
}
\label{fig:GWprodplot0}
\end{figure}

  It is also instructive to study the relative contributions to the spectrum
from GW emission during the radiation and matter eras and the effect 
of finite $t_F$.  These features are illustrated in Fig.~\ref{fig:GWprodplot0},
where we show the GW spectrum for $\alpha = 0.1$ and $G\mu = 10^{-11}$.
The solid black line shows the full spectrum, the shaded blue region
indicates the contribution from loop emission of GWs during the matter era
($t_{\rm eq} < \ttil < t_0$), and the shaded orange shows the contribution
from loop emission of GWs during the radiation era ($\ttil < t_{\rm eq}$).
Note that both the matter and radiation contributions in the figure
are dominated by loops that were formed during 
the radiation era ($t_i(t_0) < t_{eq}$).
The dashed~(dot-dashed) line in this figure shows the effect of artificially
increasing $t_F$ to the cosmological time corresponding to the radiation
temperature $T_F = 200\,\kev$~($200\,\mev$).  Finite $t_F$ imposes 
a lower cutoff on $t_i$ that modifies the spectrum when $t_i(\ttil_M) < t_F$,
implying that the spectrum falls off as $1/f$ going to large frequency.

%%%%%%%%%%%%%%%%%%%%%%%%%%%%%%%%%%%%%%%%%%%%%%%%%%%%%%%%%%%%%%%

%%%%%%%%%%%%%%%%%%%%%%%%%%%%%%%%%%%%%%%%%%%%%%%%%%%%%%%%%%%%%%%%%%%%%%

\section{Mapping the early universe with cosmic string GWs}
\label{sec:cosmomap}

  The analysis above shows the close connection between the
frequency spectrum of GWs produced by a cosmic string network and the cosmological
background when they were emitted.  In this section we investigate how 
this property could be used to test history of the very early universe 
if a relic GW signal from a string network were to be detected.  
In particular, we demonstrate that the GW spectrum could be used 
to test the standard cosmological picture further back in time than the 
best current limits based on primordial Big Bang Nucleosynthesis~(BBN).  
We also study how deviations from the standard picture
would imprint themselves on the spectrum.

\subsection{Testing the standard cosmological history}
\label{sec:cosmo_standard}
 
Current observations provide strong evidence for the standard 
$\Lambda$CDM model of cosmology~\cite{Aghanim:2018eyx}.  
In this model, the very early universe (following a period
of inflation or something similar) is dominated by radiation,
followed by a period of matter domination, and very recently entering a phase 
of accelerated expansion driven by a constant dark energy.
This evolution (after inflation) is encapsulated in the first Friedmann equation 
describing the expansion rate of the scale factor $a(t)$:
\begin{equation}\label{eq:friedmannI}
H^2 \equiv \left( \frac{\dot{a}}{a} \right)^2 = 
H_0^2\left[ 
\Delta_{R}(a)\,\Omega_R\lrf{a}{a_0}^{-4}
+ {\Omega_M}\lrf{a}{a_0}^{-3} 
+\Omega_\Lambda \right] \ ,
\end{equation} 
where $H_0 \simeq 1.44\times 10^{-42} \ {\rm GeV} $ 
is the expansion rate measured today,
$\Omega_R\simeq 9.2\times 10^{-5}$ for radiation, 
$\Omega_M \simeq 0.31$ for matter, 
and  $\Omega_\Lambda\simeq 0.69$ for dark energy~\cite{Ade:2015xua}.
The correction factor
\be \label{eq:deltarhoR}
\Delta_{R}(a)=\frac{g_*(a)}{g_*(a_0)}\left(\frac{g_{*S}(a_0)}{g_{*S}(a)}\right)^{4/3}
\ee
accounts for the deviation from $T\propto a^{-1}$ dictated by 
entropy conservation, and depends on the effective number of energy
density~($g_*$) and entropy~($g_{*S}$) degrees of freedom
for which we use the SM parametrization 
in \texttt{micrOMEGAs\_3.6.9.2}~\cite{Belanger:2014vza}.  

An early period of domination by something other than radiation would show
up in the GW frequency spectrum as a significant deviation from flatness.  
For given values of $G\mu$ and $\alpha$, the frequency $f_{\Delta}$ 
at which such a deviation would appear is determined by the cosmological 
time $t_{\Delta}$ when the (most recent) radiation 
era began.\footnote{Equivalently, radiation domination occurred 
for $t_{\Delta} < t < t_{\rm eq}$ with something else for $t < t_{\Delta}$.}  
Based on Eqs.~(\ref{eq:iscale},\,\ref{eq:fscale}) and the analysis 
in Sec.~\ref{sec:Tfrelation}, the frequency spectrum is first modified 
significantly when the dominant emission time $\ttil_M$ comes from loops created
at $t_i^{(k=1)} \simeq t_{\Delta}$.  This gives an approximate transition
frequency $f_{\Delta}$ as the solution of
\beq
t_i(\ttil_M(f_{\Delta})) = t_{\Delta} \ .
\eeq
Approximating $a(t) \propto t^{1/2}$ during the radiation era, this gives
\beq
f_{\Delta} &\simeq& 
\sqrt{\frac{8\,z_{\rm eq}}{\alpha\,\Gamma G\mu}}\,
\lrf{t_{eq}}{t_{\Delta}}^{1/2}\,t_0^{-1} \\
&\simeq&
\sqrt{\frac{8\,}{z_{\rm eq} \alpha\,\Gamma G\mu}}\,
\lrfsq{g_*(T_\Delta)}{g_*(T_0)}^{1/4}\lrf{T_{\Delta}}{T_0}\,t_0^{-1}
\nnmb
\eeq
where $z_{\rm eq} \simeq 3387$ is the redshift at matter-radiation equality,
and $T_0 = 2.725\,\text{K}$ is the temperature today.
A more accurate dependence obtained by fitting to a full numerical 
calculation that properly accounts for variations in $g_*$ gives
\be \label{eq:fdeltaforlargealpha}
f_{\Delta}= 
  (8.67\times 10^{-3} \, {\rm Hz})\, 
\lrf{T_\Delta}{\gev}
\lrf{0.1\times 50\times 10^{-11}}{\alpha\,\Gamma\,G\mu}^{1/2}
  \left(\frac{g_*(T_\Delta)}{g_*(T_0)}\right)^\frac{8}{6} \left(\frac{g_{*S}(T_0)}{g_{*S}(T_\Delta)}\right)^{-\frac{7}{6}} \ ,
\ee
which we find to be accurate to about $10\%$.

\begin{figure}[ttt]
\centering
 \includegraphics[height=7.5cm]{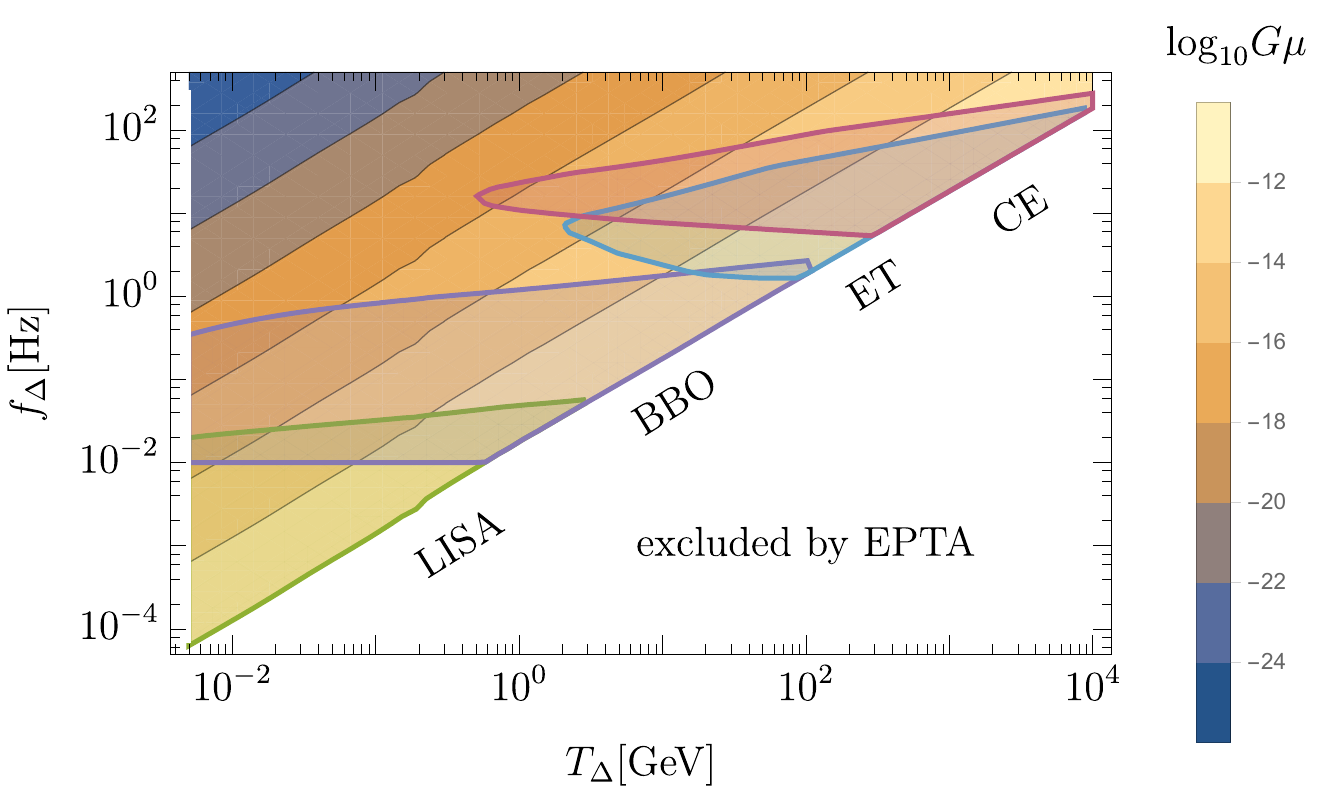}
 \caption{\label{fig:fdelta1}
Frequency $f_{\Delta}$ required to test the standard cosmology at radiation
temperature $T_\Delta$ for a range of values of $G\mu$ with $\alpha = 0.1$.
The shaded regions indicate where the signal could be detected by 
the corresponding planned future GW detector. 
}
 \end{figure}

  The power of current and future GW detectors to look back in time
using GWs from cosmic strings comes down to their sensitivity
to $f_{\Delta}$ for given values of $\alpha$ and $\Gamma G\mu$.
Measuring an approximately flat frequency spectrum out to $f_{\Delta}$
would provide strong evidence for radiation domination up to the
corresponding temperature $T_{\Delta}$.  Thus, $f_\Delta$ can be reinterpreted
as the frequency needed to test standard cosmology up to temperature $T_\Delta$.
In Fig.~\ref{fig:fdelta1} we show $f_\Delta$ as a function of $T_\Delta$ for
a range of values of $G\mu$ with $\alpha = 0.1$ and $\Gamma = 50$.
Also shown in this figure are the expected sensitivity ranges of
LISA, BBO, ET, and CE.  All four planned GW detectors could potentially
probe the standard cosmology much further back in time than BBN,
corresponding to temperatures $T_{\Delta} > 5\,\mev$. 
Note that LIGO does not appear in Fig.~\ref{fig:fdelta1} 
(and Fig.~\ref{fig:fdelta2} below) because the GW amplitude of 
the flat radiation-era plateau lies below the projected sensitivity 
of the observatory for all values of $\Gamma G\mu$ consistent with 
the pulsar timing bound of EPTA,  as can be seen in Fig.~\ref{fig:GWprodplot0}. 
However, we show in Sec.~\ref{sec:cosmo_exotic} that LIGO could be
sensitive to GW signals from cosmic strings with a non-standard 
early cosmological history.

In Fig.~\ref{fig:fdelta2} we show the expected sensitivities 
of LISA, BBO, ET, and CE in the $T_{\Delta}\!-\!G\mu$ plane.
This figure illustrates an important complementarity of the four 
detectors, corresponding to their respective ranges of frequencies.
Indeed, the ability to measure the GW signal over a broad frequency
range would be essential to establish the characteristic flat spectrum
from the radiation era.  Together, Figs.~\ref{fig:fdelta1} and \ref{fig:fdelta2}
also show that these planned observatories could probe the standard cosmological
history up to temperatures approaching $T\sim 10^4$ GeV, 
well beyond the BBN era.
 
\begin{figure}[ttt]
\centering
 \includegraphics[height=7.5cm]{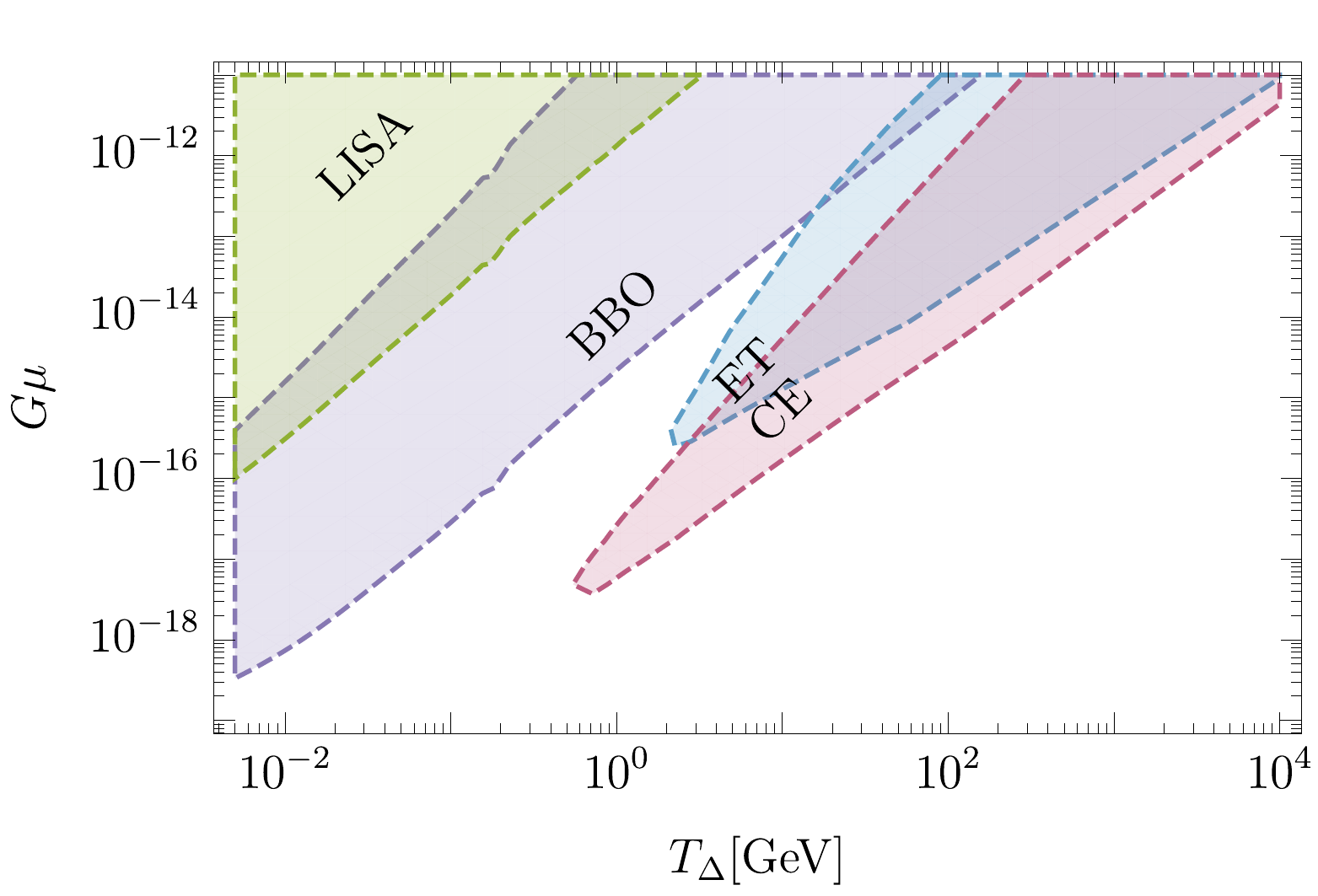}
 \caption{\label{fig:fdelta2}
Reach of future GW detectors to test the standard picture of early radiation 
domination at temperature $T_{\Delta}$ for given values 
of the string tension $G\mu$.  The shaded regions indicate where the signal 
could be detected by the corresponding planned future GW detector.
}
 \end{figure}

\subsection{Probing new degrees of freedom}\label{sec:cosmo_extra_dof}
 
  Any extension of the SM involving new particles that are thermalized
and relativistic in the early universe will contribute to the effective 
number of energy and entropy degrees of freedom~(DOFs), $g_*$ and $g_{*S}$.
Notably, the minimal supersymmetric extension of the Standard Model
predicts $g_* = 221.5$ (compared to $g_* = 106.75$ in the SM)
at temperatures above the superpartner masses and not counting the 
gravitino/goldstino.  
Other approaches to the electroweak hierarchy problem such as the 
Twin Higgs~\cite{Chacko:2005pe} or NNaturalness~\cite{Arkani-Hamed:2016rle} 
also predict many new DOFs near the weak scale.  Theories of dark matter 
with hidden sectors or hidden valleys have also attracted substantial interest 
in recent years, and frequently give rise to multiple new DOFs, 
possibly well below the weak scale~\cite{Kolb:1985bf,Hodges:1993yb,Strassler:2006im,Feng:2008mu,Foot:2014mia,Adshead:2016xxj}.  
In some cases, the increase in $g_*$ can be enormous~\cite{Kaplan:2015fuy,Giudice:2016yja,Soni:2016gzf}.  

  New DOFs are already constrained by cosmological observations for masses
below a few MeV.  Very light states ($m\lesssim \mathrm{eV}$) increase
the radiation density at recombination, leaving an imprint on the CMB.
This effect is usually expressed as an equivalent number of additional
neutrino species \cite{Brust:2013xpv,Chacko:2015noa,Baumann:2015rya,Brust:2017nmv, Cui:2018imi}, with measurements limiting 
$\Delta N_{eff}^{CMB} < 0.30$~\cite{Aghanim:2018eyx}.
More massive states can avoid the CMB bound, but can modify the expansion
rate during BBN, with the limit for masses below $m\lesssim \mev$ given by
$\Delta N_{eff}^{BBN} \lesssim 0.5$~\cite{Cyburt:2015mya}.  
In this section we study the effect of new DOFs on the GW spectrum 
from cosmic strings.  
We show that detailed GW frequency measurements could probe new, more massive
DOFs beyond what can be inferred from the CMB or BBN.  Earlier suggestions 
along this line with a focus on SM degrees of freedom 
can be found in Refs.~\cite{Battye:1997ji,Saikawa:2018rcs}. 
  
\begin{figure}
\centering
\includegraphics[height=6.5cm]{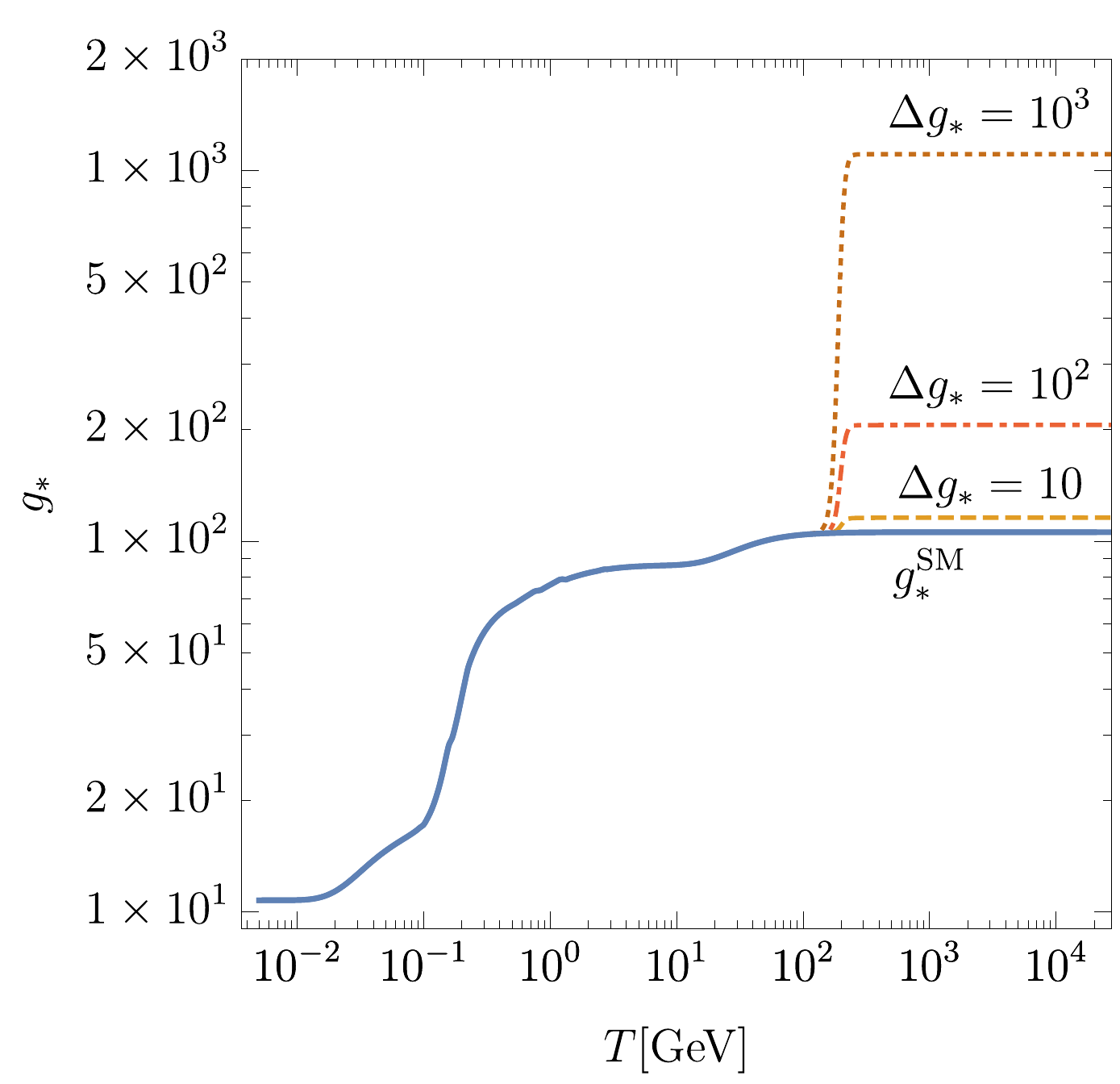}
\includegraphics[height=6.5cm]{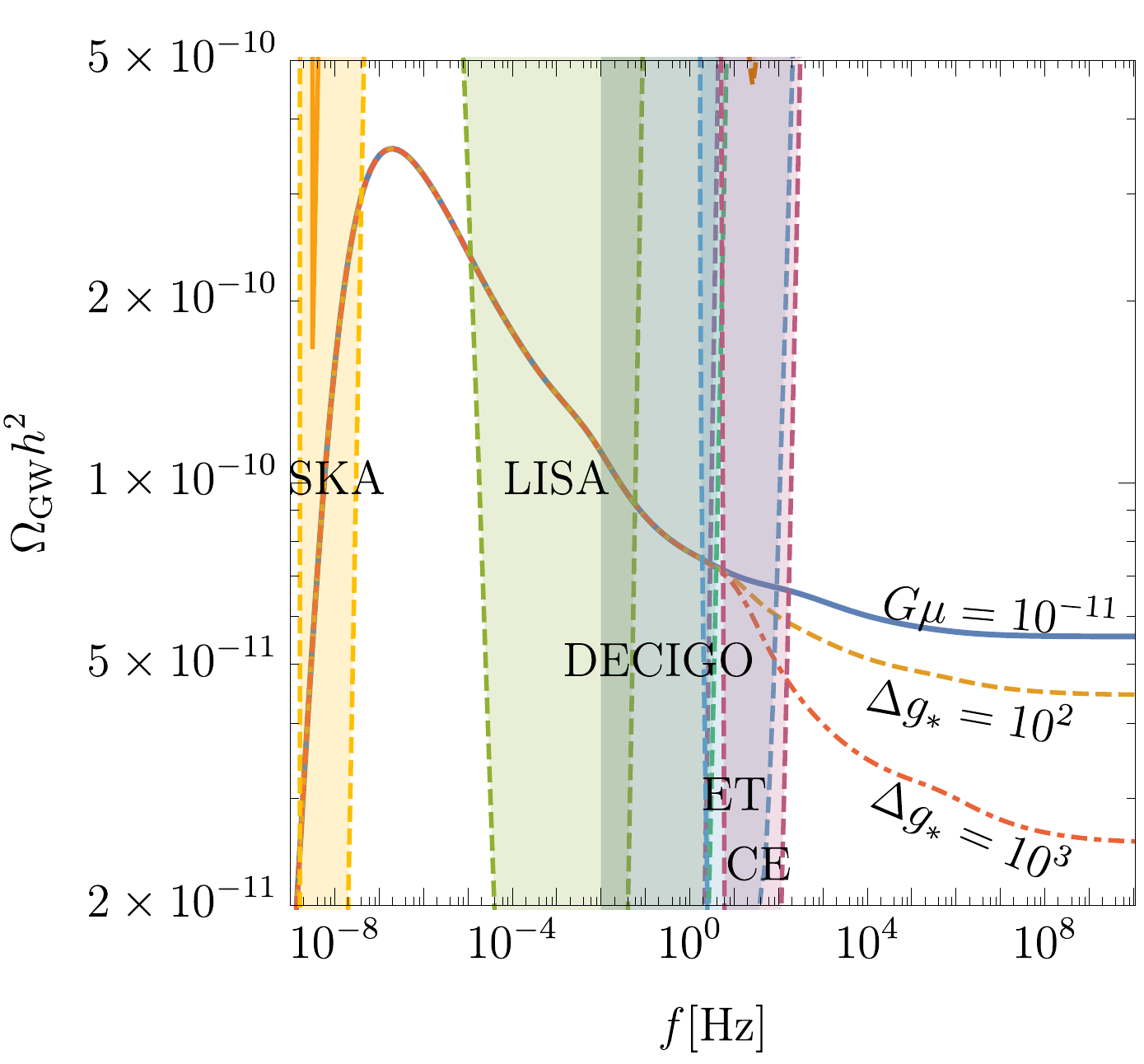}
\caption{
Left panel: illustration of our parametrization of the change in the number of DOFs for $T_\Delta=200\,\gev$ and $\Delta g_* = 10^1,\,10^2,\,10^3$. Right panel: modification of the GW spectrum from a cosmic string network with $G\mu = 10^{-11}$ and $\alpha = 0.1$ for this modification in $g_*$.  The colored regions in this panel show the expected sensitivities of SKA, LISA, DECIGO, ET, and CE.
\label{fig:dofexample}}
\end{figure}

  To illustrate the generic effect of new massive DOFs on the string GW spectrum
without reference to a specific extension of the SM, 
we model the change by a rapid decrease in $g_*$ as the temperature falls 
below the mass threshold $T_\Delta$:
\be \label{eq:dofmod}
g_*(T)=
g_*^{\rm SM}(T) + \frac{\Delta g_*}{2}
\left[1+\tanh\left(10\,\frac{T-T_{\Delta}}{T_{\Delta}} \right)\right]\approx
\begin{cases}
g_*^{\rm SM}(T)  \quad \quad \quad \quad {\rm ;} \quad T<T_{\Delta}
\\ 
g_*^{\rm SM}(T) + \Delta g_* \quad {\rm ;} \quad T>T_{\Delta}
\end{cases} \ .
\ee
An identical modification is assumed for $g_{*S}$, and we use entropy
conservation to derive the temperature dependence on the scale factor
through the decoupling transition.  The resulting dependence of 
$g_* = g_{*S}$ on $T$ is shown in the left panel of Fig.~\ref{fig:dofexample}
for $\Delta g_* = 10^1,\,10^2,\,10^3$ at $T_\Delta = 200\,\gev$.  

  In the right panel of Fig.~\ref{fig:dofexample} we show the effect
of changing $g_*$ on the GW spectrum from a cosmic string network
with $G\mu = 10^{-11}$ and $\alpha=0.1$, again for 
$\Delta g_* = 10^1,\,10^2,\,10^3$ at $T_\Delta = 200\,\gev$. 
The shaded regions in this panel show the estimated sensitivity bands
of SKA, LISA, DECIGO, ET, and CE as in previous figures.  A fractional change in
$g_*$ by order unity or more is seen to produce a significant and potentially
observable decrease in the cosmic string GW amplitude above a specific frequency.
This transition frequency $f_{\Delta}$ is determined by $T_\Delta$
but is independent of $\Delta g_*$.  For $f \gg f_{\Delta}$, 
the GW spectrum returns to a flat plateau characteristic of 
radiation domination~(RD) but with a smaller amplitude.
The result of Fig.~\ref{fig:dofexample} also shows that future GW detectors
could be sensitive to new DOFs with masses relevant
to solutions to the electroweak hierarchy problem,
possibly even beyond the reach of the LHC.
We have checked that this result is insensitive to the precise form 
of the interpolation function used for $g_*$, relative to Eq.~\eqref{eq:dofmod}, 
as long as it varies reasonably quickly.

  The change in the spectrum shown in the right panel of Fig.~\ref{fig:dofexample}
can be understood in terms of the frequency--temperature correspondence derived
in Sec.~\ref{sec:Tfrelation}.  As expected, the GW spectrum is only modified 
above the transition frequency $f_{\Delta}$, which is determined by the 
temperature (time) at which the standard cosmology is modified.  
In contrast to this analysis, however, the change in the cosmological 
evolution from massive decoupling is more subtle than a change 
in the dilution exponent of the energy density.  Even so, a simple analytic
estimate of the change in the amplitude is possible.  

Since the main contribution to the amplitude at high frequencies is
expected to come from deep in the RD era, the Hubble rate and time
for large $T$ can be approximated by
\beq
H \approx \sqrt{\Delta_R \Omega_R}\,H_0\,a^{-2}
\ , \quad \quad
 t \approx \frac{a^2}{2\sqrt{\Delta_R \Omega_R}} \ .
\eeq
With this simplification, the integral in Eq.~\eqref{eq:GWdensity2} 
can be written directly in terms of the scale factor to give
\be
\label{eq:OmegaPlateau}
\Omega_{\rm GW}(f) \approx \frac{128\pi}{9} \Delta_R\,\Omega_R\,
\frac{C_{\rm eff}(n\!=\!4)}{\Gamma}\;\alpha^{1/2}\,
(\Gamma G\mu)^{1/2} \ ,
%( \sqrt{\frac{G\mu \alpha}{\Gamma}} \ ,
\ee
which agrees well with a similar calculation 
in Ref.~\cite{Blanco-Pillado:2017oxo}.
This implies that the amplitude of the RD plateau depends on the 
number of DOFs via $\Delta_R$, and thus
\be
\Omega_{\rm GW}(f\gg f_\Delta) \approx
\Omega_{\rm GW}^{\rm SM}(f)
\left(  \frac{g_*^{\rm SM}}{g_*^{\rm SM}+\Delta g_*} \right)^{1/3}, 
\label{eq:gstar_effect}
\ee
where $\Omega_{\rm GW}^{\rm SM}$ is the amplitude with only SM DOFs,
and we have assumed $g_* = g_{*S}$ at high $T$.
Therefore an increase of number of DOFs at $T_\Delta$ leads to a drop 
in the amplitude at frequencies above $f_\Delta$.  
In fact, similar changes in the GW amplitude from the RD era from changes
in the number of effective SM degrees of freedom at the QCD phase transition
and electron-positron decoupling are visible in Figs.~\ref{fig:GWGmuplot} 
and \ref{fig:GWprodplot0}.  
We also find that the magnitude of the amplitude decrease 
in Eq.~\eqref{eq:gstar_effect} agrees well with the full numerical 
result shown in the right panel of Fig.~\ref{fig:dofexample}.

%%%%%%%%%%%%%%%%%%%%%%%%%%%%%%%%%%%%%%%%%%%%%%%%%%%%%%%%%%%%%%%%%%%%%%%%%%%%%%%%%%%%%%%%%%%%%%%
%%%%%%%%%%%%%%%%%%%%%%%%%%%%%%%%%%%%%%%%%%%%%%%%%%%%%%%%%%%%%%%%%%%%%%%%%%%%%%%%%%%%%%%%%%%%%%

\subsection{Probing 
new phases of cosmological evolution
%new sources of cosmological energy density
\label{sec:cosmo_exotic}
}

The second type of cosmological modification we consider is an early period
in which the expansion of the universe is driven by a new source
of energy density prior to the most recent radiation era, 
leading to a non-standard equation of state in the early universe.  
For example, an early epoch of matter domination with $\rho \propto a^{-3}$ 
can arise from a large density of a long-lived massive
particle or oscillations of a scalar moduli field in a quadratic 
potential~\cite{Moroi:1999zb}. Such a period of matter domination ends when 
the long-lived species decays to the SM.  A more exotic class of deviations can
arise from the energy density of a scalar field $\phi$ oscillating in a potential
of the form $V(\phi) \propto \phi^N$, which gives $n = 6N/(N+2)$.  
In the extreme limit of $N\to \infty$ we have $n\to 6$, corresponding to 
the oscillation energy being dominated by the kinetic energy of the scalar.
This behavior arises in models of inflation, quintessence, dark energy, 
and axions, and is called kination~\cite{Salati:2002md,Chung:2007vz, Poulin:2018dzj}.  For all these cases, the universe must settle to radiation domination 
by the time the temperature reaches $T_{\Delta} \sim 5\mev$ in order 
to preserve the successful predictions of BBN~\cite{Hannestad:2004px}.
    
  To model the effect of a new cosmological energy source, we parametrize 
the evolution of the energy density of the universe according to
\begin{equation} \label{eq:energydensityevolution}
\rho(t) =
\begin{cases}
\rho_{st}(t_{\Delta})\left[\frac{a(t)}{a(t_{\Delta})}\right]^{-n}
 & \ {;} \ \ t < t_{\Delta}
\\
\rho_{st}(t) & \ {;} \ \ t \geq t_{\Delta}
\end{cases}
\end{equation}
where $\rho_{st}(t)$ is the standard energy density 
given by Eq.~\eqref{eq:friedmannI}.  In this context, we define
$T_{\Delta}$ as the radiation temperature at time $t_\Delta$
when the recent period of radiation domination begins.
We also focus on the specific cases of $n=3$ and $n=6$ since these bound 
the envelope of the set of well-motivated possibilities discussed above.  

\begin{figure}[ttt]
\centering
 \includegraphics[height=6.8cm]{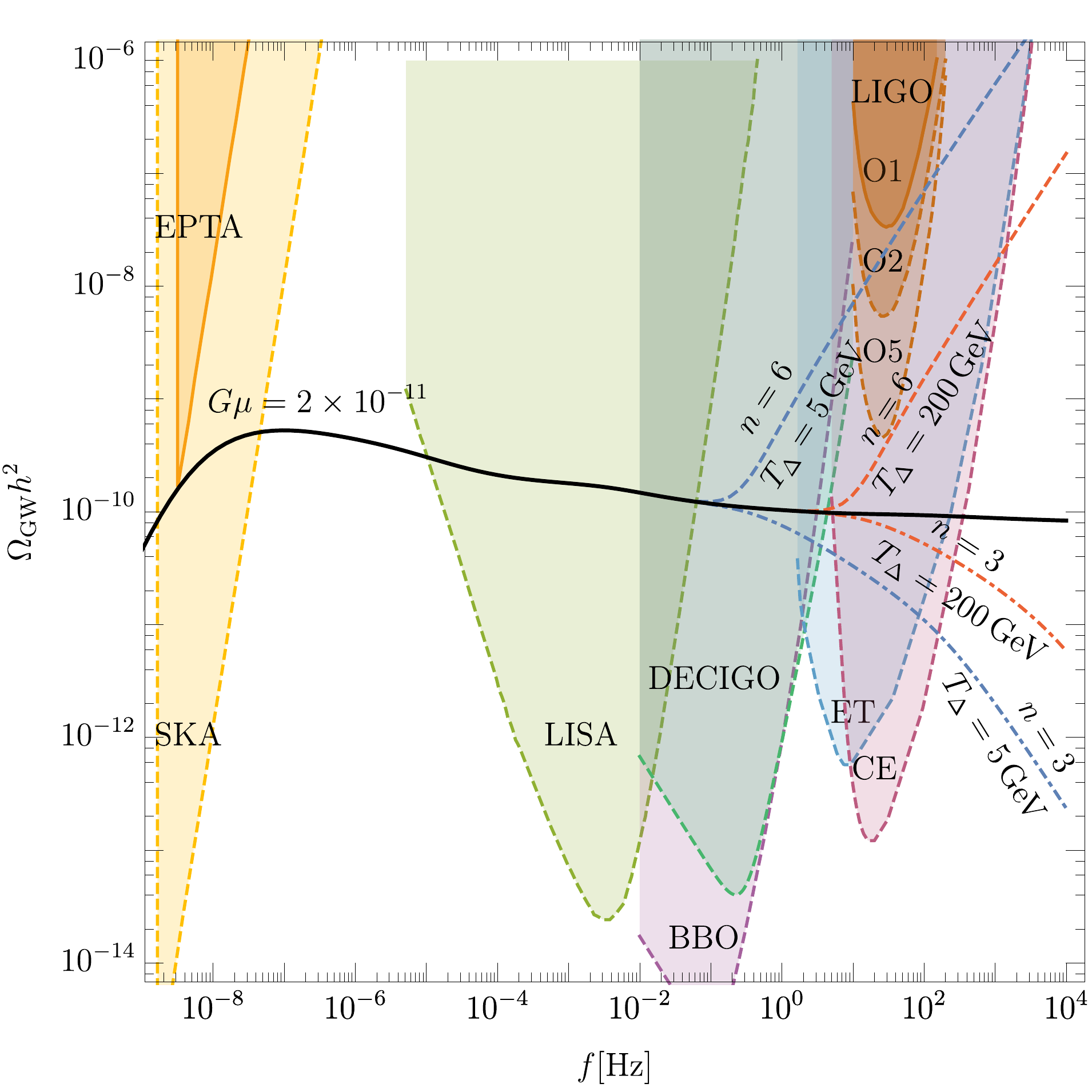}
  \includegraphics[height=6.8cm]{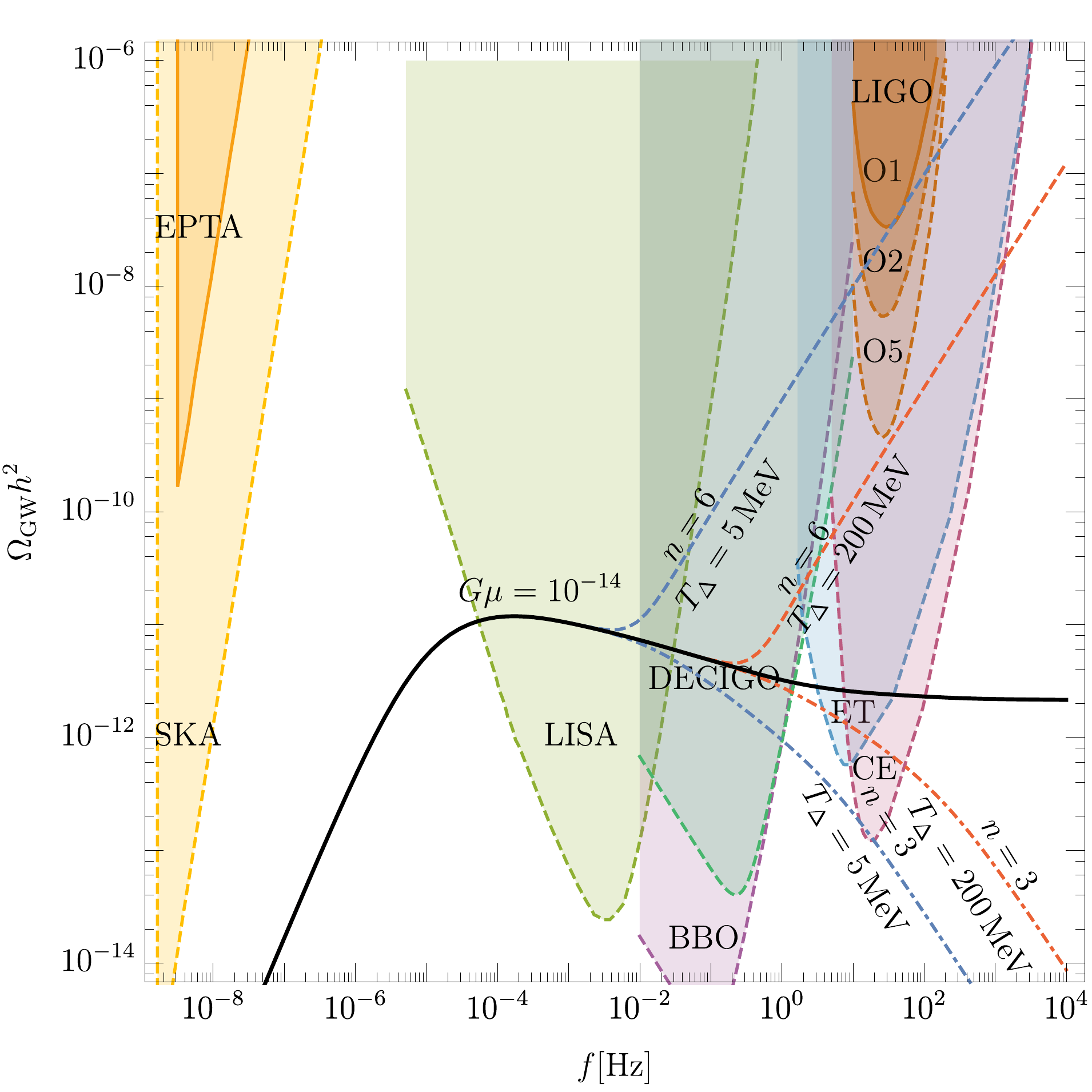}
 \caption{\label{fig:gwmod}
Frequency spectra of gravitational waves $\Omega_{\rm GW}h^2$ sourced by a
scaling cosmic string networks with $G\mu = 2 \times 10^{-11}$~(left) 
and $10^{-14}$~(right) with $\alpha=10^{-1}$.
The solid black lines show the spectra for a standard cosmological evolution, 
while the dashed~(dash-dotted) lines indicate the result with in an early 
period of $n=6$ kination~($n=3$ matter) domination lasting until
the temperatures $T_{\Delta}$ listed in the figure.  Also shown are the 
sensitivity bands of current and future GW detectors.
} 
 \end{figure}

In Fig.~\ref{fig:gwmod} we show the GW spectra from 
cosmic strings for $\alpha = 0.1$ and $G\mu = 2\times 10^{-11}$~(left)
and $G\mu = 10^{-14}$~(right), together with the modifications to the spectra for
early periods of domination with $n=3$ or $n=6$ at representative transition
temperatures $T_\Delta$.  For $G\mu = 2\times 10^{-11}$ on the left we show
$T_\Delta = 5\gev$ and $200\gev$, and  for $G\mu = 10^{-14}$ on the right we take
$T_{\Delta} = 5\mev$ and $200\mev$.  For reference, we also indicate the expected 
sensitivities of current and future GW detectors. 

  The onset and shape of the modifications to the GW spectra can be
understood in terms of our previous analytical estimates.
In particular, Eq.~\eqref{eq:fdeltaforlargealpha} gives a good approximation
of the frequency $f_{\Delta}$ above which the spectrum deviates significantly
from the standard cosmology, while Eq.~\eqref{eq:fscale} 
describes the frequency dependence beyond this.  
Applying Eq.~\eqref{eq:fscale}, we find 
$\Omega_{GW}(f> f_\Delta) \propto f^{+1}\,(f^{-1})$ for $n=6\,(3)$.
The modifications to the spectrum from the flat plateau of the standard early
RD era are drastic and observable provided they occur at low enough frequency 
to fall within the sensitivity range of current or future experiments. 

  Relative to the standard cosmology, we also note that an early phase 
with $n > 4$ tends to be easier to observe because it implies a rising
amplitude at high frequency.  
Correspondingly, the experimental sensitivities
indicated in Figs.~\ref{fig:fdelta1} and \ref{fig:fdelta2} are lower bounds 
on what can be tested for modified cosmologies with $n> 4$.
Moreover, the left panel of Fig.~\ref{fig:gwmod} shows that future phases
of LIGO could probe an $n=6$ modified cosmology up to 
the transition temperature $T_\Delta = 200\,\gev$ for $G\mu=2\times 10^{-11}$,
which is about as large a $G\mu$ as possible given current limits 
from pulsar timing.  This range could be extended even further by 
the proposed ET and CT observatories.  
However, let us also mention that the maximal GW amplitude is constrained 
by the total radiation density in GWs, 
corresponding to~\cite{Smith:2006nka,Henrot-Versille:2014jua}
\beq
\int\!d(\ln f)\,\Omega_{GW} \lesssim 3.8\times 10^{-6} \ .
\eeq
This bound limits the duration of an early phase with $n>4$.

%%%%%%%%%%%%%%%%%%%%%%%%%%%%%%%%%%%%%%%%%%%%
%%%%%%%%%%%%%%%%%%%%%%%%%%%%%%%%%%%%%%%%%%%%

\section{Detection challenges and ways to overcome them}
\label{sec:detect}

Up to now we have only studied whether the GW signals from a cosmic string
network lie within the sensitivity reach of current and future detectors.
In this section we confront the practical challenges of subtracting 
astrophysical backgrounds and  identifying whether such a signal is due
to cosmic strings or some other phenomenon. We also examine other potential 
sources of GW signals and the extent to which they can be distinguished 
from cosmic strings evolving in a standard or non-standard cosmological 
background.  Finally, we comment on how smaller values of the initial loop size 
parameter $\alpha$ would impact our results.

\subsection{Astrophysical backgrounds}

  The LIGO/Virgo experiment has already observed a number of binary mergers
of black holes and neutron stars~\cite{TheLIGOScientific:2017qsa,Abbott:2017gyy,Abbott:2017vtc,TheLIGOScientific:2016wyq}.  Based on the number of mergers seen 
and assuming the redshift dependence of the merger rate follows that of 
star formation, it is anticipated that Advanced LIGO/Virgo will soon begin 
to detect a stochastic GW background from a collection of weaker, 
unresolved binary mergers~\cite{Abbott:2017xzg,TheLIGOScientific:2016dpb}.  
This background is expected to begin and peak near $f \sim 1000\,\mathrm{Hz}$ 
with $\Omega_{\rm GW} \sim 10^{-9}$, and fall off in amplitude as $f^{2/3}$ 
at lower frequencies~\cite{Moore:2014lga,Zhu:2012xw}, also
putting it within the detection range of space-based detectors such as LISA.
The signal from these unresolved mergers will also overlap with, and sometimes
overwhelm, the prediction for cosmic strings.

  For larger values of $G\mu \gtrsim 10^{-15}$, the lower frequency portion 
of the cosmic string GW signal could be observed in space-based detectors
well above the expected background from binary mergers.  This is likely to
include a part of the characteristic flat portion of the spectrum from GW emission
in the radiation era.  In contrast, the higher frequency portion of the
cosmic string signal from a standard cosmological history in the sensitivity 
range of LIGO is likely to be swamped by the binary background.
However, let us point out that an enhanced cosmic string signal 
due to non-standard cosmology with an early phase of $n > 4$ evolution 
could potentially rise above background.  We also note that the $f^{2/3}$ 
rise in the binary background spectrum has the same frequency scaling as
the cosmic string spectrum with an early period of $n=5$ evolution.

  Significant effort has been put into finding ways to subtract the
background from binary mergers~\cite{Harms:2008xv,Yagi:2011wg,Regimbau:2016ike,Caprini:2018mtu,Jenkins:2018nty,Raidal:2017mfl,Carr:2017jsz,Carr:2016drx,Smith:2017vfk}.  
Since the stochastic background from binary mergers comes from those
that are not resolved, a promising strategy is to use the improved
angular sensitivity of future detectors to identify a great number of them,
thereby reducing the portion that contribute to the effective stochastic signal~\cite{Regimbau:2016ike,Caprini:2018mtu,Jenkins:2018nty}. 
In particular, Ref.~\cite{Regimbau:2016ike} suggests that these backgrounds 
can be removed to the level of $\Omega_{\rm GW} \sim 10^{-13}$ in future 
ground-based detector arrays such at ET and CE, 
while \cite{Harms:2008xv,Yagi:2011wg} find even better sensitivity 
for BBO after background subtraction.
For LIGO, a statistically optimal search strategy has been proposed recently
for identifying unresolved binaries that offers a significant improvement 
relative to using the traditional cross-correlation method~\cite{Smith:2017vfk}.
These studies suggest that the cosmic string GW signals discussed in this work 
can be separated over background to an extent that they remain a powerful
tool to probe the early universe.

\subsection{Distinguishing cosmic strings from other new phenomena}

  Cosmic strings are just one of many forms of new physics that can give
rise to stochastic GW signals~\cite{Binetruy:2012ze,Caprini:2018mtu,Kuroyanagi:2018csn}.  
Other possibilities include primordial 
inflation~\cite{Grishchuk:1974ny,Starobinsky:1979ty,Allen:1987bk},
preheating~\cite{Khlebnikov:1997di,Easther:2006gt,Easther:2006vd,GarciaBellido:2007dg}, 
first-order phase transitions~\cite{Witten:1984rs,Hogan:1986qda,Kosowsky:1991ua}, 
and other types of topological defects~\cite{Gleiser:1998na,Hiramatsu:2013qaa}.  
Should a new (non-astrophysical) GW signal be observed, 
identifying the nature of its source will be of paramount importance.
Furthermore, if a signal due to cosmic strings is to be used to test the
cosmological history of the universe, it must be distinguished from
other types of new physics.

  A characteristic feature of the GW frequency spectrum from a scaling cosmic
string network is the flat plateau at higher frequencies.  This feature is
difficult to reproduce by most other new sources of GWs.  For example,
the GW signals from strongly first-order cosmological phase transitions
have been studied extensively~\cite{Caprini:2015zlo,Weir:2017wfa,Cai:2017cbj},
often in connection with electroweak symmetry breaking or baryogenesis~\cite{Grojean:2006bp,Caprini:2009fx,Ashoorioon:2009nf,Chung:2010cb,Chala:2016ykx,Huang:2016cjm,Beniwal:2017eik}.
%\cite{Vaskonen:2016yiu,Ashoorioon:2009nf,Enqvist:2014zqa,Artymowski:2016tme,Tenkanen:2016idg,Huang:2016cjm,Kakizaki:2015wua,Hashino:2016rvx,Hashino:2016xoj,Kobakhidze:2016mch,Chala:2016ykx,Choi:1993cv,Beniwal:2017eik,Kang:2017mkl,Kobakhidze:2017mru,Cai:2017tmh}.
The resulting spectrum typically increases following a power law 
in frequency with a positive exponent up to a peak, and then falls as 
a power law with a negative exponent at higher frequencies.
Other new sources of GWs typically also display such power-law frequency 
dependence~\cite{Binetruy:2012ze,Caprini:2018mtu,Kuroyanagi:2018csn}.
The partially flat spectrum from a cosmic string network (with standard
cosmology) can be distinguished from such a rising and falling
spectrum provided the signal can be measured over a reasonably 
broad frequency range.  We emphasize that multiple detectors may 
be needed to do so.  Separating the cosmic string spectrum with an early
phase of $n< 4$ from the spectrum due to a phase transition would be
more challenging, especially if the transition temperature is relatively low.  
However, even such more complicated scenarios could potentially be identified
through precise measurements of the frequency dependence.

  A notable exception to the typical split power-law spectrum of
new GW sources is the GW signal created by minimal models of inflation.
If the inflationary power spectrum is nearly scale invariant, the GW background
is expected to be flat over many decades in frequency corresponding to 
frequencies that reenter the horizon during radiation 
domination~\cite{Turner:1993vb,Allen:1996vm}. The stochastic spectrum
rises as $f^{-2}$ at lower frequencies ($f\lesssim 10^{-16}\,\mathrm{Hz}$) 
correspondng to modes that entered the horizon in 
the matter era~\cite{Turner:1993vb,Allen:1996vm}.  
This is analagous to the flat GW spectrum from cosmic strings during
radiation and the rise as $f^{-1/2}$ related to the matter era.
However, the amplitude inflationary GW spectrum is severely constrained 
by CMB isotropy and polarization measurements ($f \sim 10^{-17}\,\mathrm{Hz}$) 
which constrain the amplitude of the flat part of the spectrum to be less 
than about $\Omega_{\rm GW} \lesssim 10^{-15}$~\cite{Lasky:2015lej,Smith:2005mm}.
A partially flat spectrum due to cosmic strings would then be identifiable
simply through its larger amplitude, and possibly by its different power law
dependence at lower frequencies.  Note that non-minimal models 
of inflation~\cite{Barnaby:2011qe},
reheating effects~\cite{Kuroyanagi:2014nba}, or non-standard neutrino
interactions~\cite{Ghosh:2017jdy} could potentially create a larger signal 
that rises with frequency.
An early phase of kination may also increase the signal amplitude 
at higher frequencies~\cite{Giovannini:1998bp,Riazuelo:2000fc,Sahni:2001qp,Tashiro:2003qp}.

\subsection{Sensitivity to the loop size parameter $\alpha$}
\label{sec:alphasensitivity}

Recent simulations of cosmic string networks find a population of large
loops with initial loop size parameter peaked near $\alpha \simeq 0.1$~\cite{Blanco-Pillado:2013qja,Blanco-Pillado:2017oxo}.  We have used this as a fiducial
value throughout the work.  However, there is some uncertainty in the peak value
as well as the distribution around it.  Since the amplitude and frequency
dependence of the cosmic string GW spectrum depend on $\alpha$, 
this represents a further challenge to identifying the nature of the 
early universe through the spectrum.

\begin{figure}
\centering
\includegraphics[height=8cm]{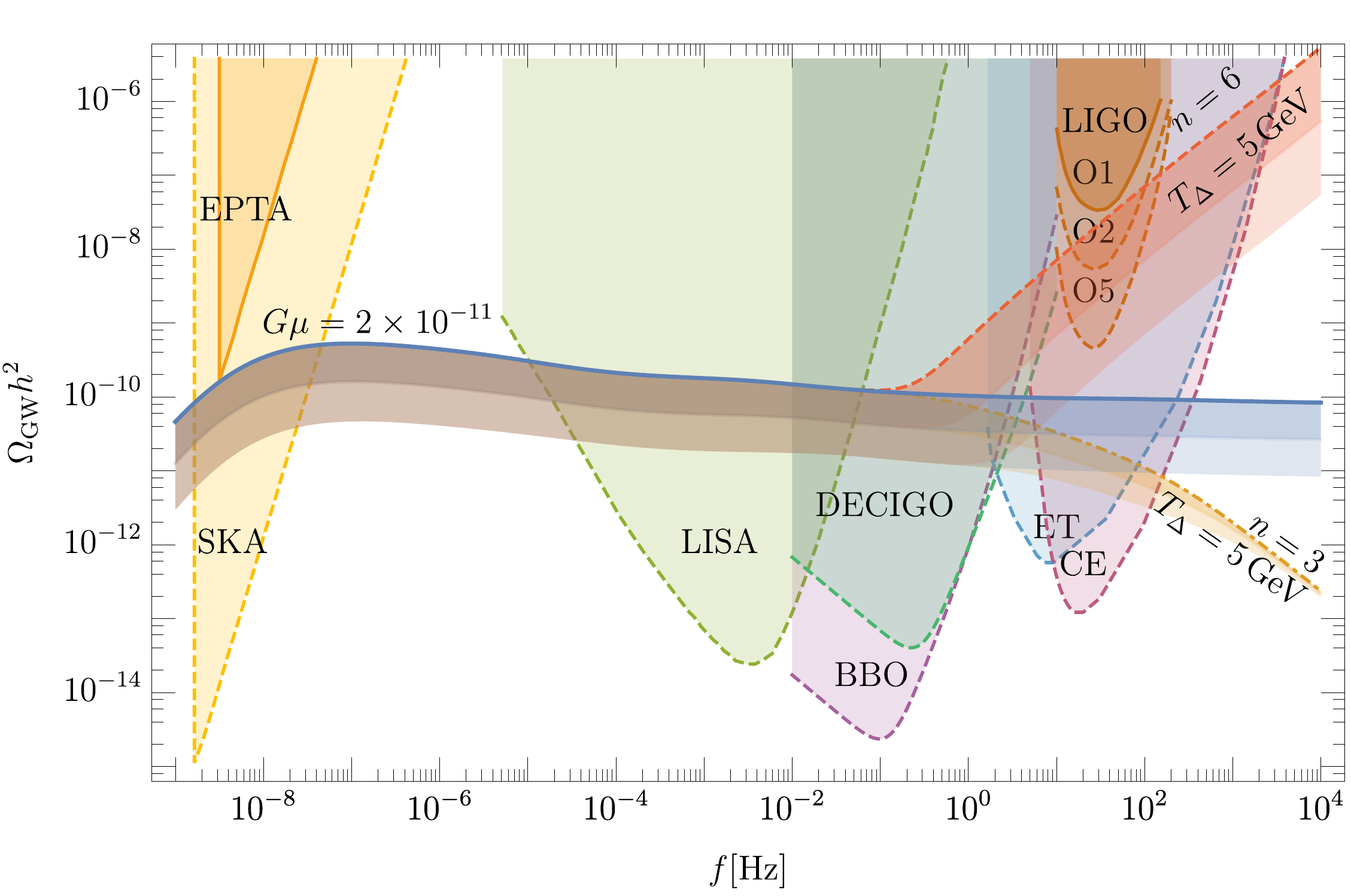}
\caption{Gravitational wave spectra from a cosmic string network with 
$G\mu =2\times 10^{-11}$ and $\alpha = 10^{-3}\!-\!10^{-1}$.  
The solid blue line shows the spectrum for the standard cosmological history 
with $\alpha = 0.1$ while the dark~(light) blue band around it shows the effect
of reducing to $\alpha = 10^{-2}$~($10^{-3}$). 
The red dashed~(orange dash-dotted) lines indicate the result for $\alpha=0.1$ 
with in an early period of $n=6$ kination~($n=3$ matter) domination 
lasting down to temperature $T_{\Delta} = 5\,\mev$.
As before, the dark~(light) colored bands around the lines show the effects 
of reducing the initial loop size to $\alpha = 10^{-2}$~($10^{-3}$).
}
\label{fig:AlphaUncertaintyPlot}
\end{figure}

In Fig.~\ref{fig:AlphaUncertaintyPlot} we show the cosmic string GW spectrum
for $G\mu = 2\times 10^{-11}$ while varying the loop size parameter
over the range $\alpha = 10^{-3}\!-\!10^{-1}$.
The solid blue line shows the spectrum for the standard cosmological history 
with $\alpha = 0.1$ while the blue band around it shows the effect of
reducing this parameter to $\alpha = 10^{-2}$~(dark blue band) and 
$\alpha = 10^{-3}$~(light blue band).  Similarly, the red 
dashed~(orange dash-dotted) lines indicate the result for $\alpha=0.1$ 
with in an early period of $n=6$ kination~($n=3$ matter) domination 
down to temperature $T_{\Delta}=5\, {\rm GeV}$.  Again, the shaded bands
show the effects of reducing $\alpha$ down to $10^{-2}$ and $10^{-3}$.

The dependence on $\alpha$ for the standard cosmological history 
shown in Fig.~\ref{fig:AlphaUncertaintyPlot} matches our previous analysis 
in Sec.~\ref{sec:cosmomap}, with the amplitude of the radiation-era plateau 
varying as $\Omega \propto (\alpha\,\Gamma G\mu)^{{1}/{2}}$ 
(Eq.~\eqref{eq:OmegaPlateau}).
Furthermore, the frequency at which the spectrum is first modified 
by non-standard cosmology varies as 
$f_{\Delta} \propto (\alpha\,\Gamma G\mu)^{-1/2}$ 
(Eq.~\eqref{eq:fdeltaforlargealpha}).
For early matter domination ($n=3$) the modifications 
to the frequency and amplitude cancel out, leaving a falling slope 
at high frequency unchanged.  In contrast, for early kination ($n=6$)
the changes to the frequency and amplitude add to give a linear relation
$\Omega_{GW}\propto \alpha^1$.  While an uncertainty in $\alpha$ would
complicate the identification of a transition temperature $T_\Delta$,
it does not make it impossible.  In principle, the combination
$\alpha\,\Gamma G\mu$ could be extracted from the amplitude of a flat
radiation plateau and then applied to obtain $T_\Delta$ from an
observation of $f_\Delta$ via Eq.~\eqref{eq:fdeltaforlargealpha}.

%%%%%%%%%%%%%%%%%%%%%%%%%%%%%%%%%%%%%%%%%%%%%%%%%%%%%%%%%%%%%%%

%======================================================
\section{Conclusions} \label{sec:conc}

Standard cosmology maintains that an era of radiation domination began in the early universe and was followed by matter domination, which then ultimately yields to an increasing acceleration era dominated by the cosmological constant. This framework is well tested and is found to be self-consistent by a multitude of experimental probes including measurements of the CMB, supernovae, large-scale structure, and abundances of nuclei as predicted by BBN epoch. 

Unfortunately, the traditional experimental probes reach back only as far as BBN, which corresponds to temperatures below only about 5 MeV. There are many ideas for new physics above 5 MeV that disrupt the standard cosmology, whether it be through a different scaling phase other than radiation domination (e.g., matter or kination domination), or through extra degrees of freedom beyond the known Standard Model ones that  substantially modify radiation era dynamics. Therefore, testing for new physics, and an altered cosmological evolution at temperatures greater than $5\mev$, requires new methods. The potential answer is gravitational waves, whose very early origins pass safely through recombination and BBN, which scrambles the otherwise powerful CMB probes and BBN constraints.

A strong early universe source of GWs must be present in order to probe the effects that cosmological evolution can have on it. Furthermore, this source must have a reasonably well understood emission spectrum -- analogous to the \textit{standard candles} of supernovae -- with which to propagate through various assumed cosmological histories and compare with observational data. A prime candidate for this is cosmic strings, whose network formation and emission spectrum has been well studied and understood, particularly featuring a long flat plateau at high frequency during standard radiation dominated era. Another reason cosmic stings are useful GW sources to consider is that they are generically expected in a wide variety of high-scale theories of particle physics, ranging from unified field theories containing abelian factors to fundamental string theory.

We have assumed the existence of cosmic strings in the early universe and have worked out the GW relic abundance vs.\ frequency spectrum for many different cosmic string tensions $G\mu$. We have reiterated previous results in the literature that GWs are an excellent way to constrain and find evidence for cosmic strings even within standard cosmological evolution (see Fig.~\ref{fig:GWGmuplot}). In addition, and what is central to our study, the GWs from cosmic strings enable the probing of modifications of early universe cosmology in regimes that no other probe can.

We studied two main ways that early universe cosmology can change. First, we studied the effect of having a very large number of additional degrees of freedom present in the spectrum at high energy. If the degrees of freedom are present down to temperature $T_\Delta$ one finds that there is a  frequency  $f_\Delta$ above which the GW energy density is altered compared to the expectations of standard cosmology (with SM degrees of freedom). The signal for the onset of a high number of degrees of freedom is therefore standard $\Omega_{\rm GW}(f)$ vs.\ $f$ for cosmic strings up to $f_\Delta$ and then a fall-off for $f>f_\Delta$ compared to expectations. Fig.~\ref{fig:dofexample} shows the effect in the $\Omega_{\rm GW}(f)$ vs.\ $f$ plane.

A second example is GWs from cosmic strings evolving in a non-standard phase, either of an early matter domination phase $(n=3)$ or an early kination $(n=6)$ phase. The early matter phase may be due the presence of a large density of heavy new physics states that later decay bringing the universe back to radiation era, which is needed to satisfy BBN constraints. In other words, the universe transitions from radiation domination at very high temperatures to matter domination (at $T$ comparable to mass of long-lived heavy new particles) and then back to radiation domination (by decay of said particles) before the onset of BBN. The kination ($n=6$) phase arises from oscillating scalar moduli in the early universe, which then decay. This leads to a cosmological history of very early radiation domination to kination domination (oscillation energy dominating) and back to radiation (by decay of the moduli).

The ability to probe these alternative cosmological histories well by cosmic strings partly derives from the property that cosmic strings rapidly enter a scaling regime, which means their energy density scales with scale factor $a$ exactly the same as the dominant energy density of the universe. If there is an early matter domination phase then GW energy density scales like $a^3$ during that phase, and if there is an early kination phase, cosmic strings will scale like $a^6$ during that phase. The scaling behavior of cosmic strings means that the energy density of the GWs emitted will be altered substantially through its non-standard redshifting. Our numerical work shows the effect quantitatively, which leads to a sharp fall-off in $\Omega_{\rm GW}(f)$ at high frequency $f$ (corresponding to the new phase era) if there is early matter domination, and a sharp rise in $\Omega_{\rm GW}(f)$ if there is an early kination phase. The results are illustrated in Fig.~\ref{fig:gwmod}.

GW detectors have given us a window to early universe cosmology complementary to any other probes previously developed. We have argued that a strong and well-understood source of GWs in the early universe could give us unprecedented ability to probe cosmological energy evolution of the early universe far earlier than previously attainable. We have also demonstrated that cosmic strings, if they exist, would be excellent standard candles to achieve these aims.

%GW detectors have given us a window to early universe cosmology complementary to any other probes previously developed. We have argued that any strong and well-understood source of GWs in the early universe will give us unprecedented ability to probe cosmological energy evolution of the early universe far earlier than previously attainable. We have also argued that cosmic strings, if they exist, would be excellent standard candles to achieve these aims.

\section*{Acknowledgments}
We thank Jose J. Blanco-Pillado, Nikita Blinov, Bhaskar Dutta, Vuk Mandic, 
and David McKeen for helpful discussions.
ML and YC would like to thank the Mainz Institute for Theoretical Physics (MITP) 
for its hospitality and support.
YC is supported in part by the US Department of Energy~(DOE) grant DE-SC0008541.
ML is supported by the United Kingdom STFC Grant ST/P000258/1 
and by the Polish MNiSW grant IP2015 043174.
DEM is supported by a Discovery Grant from the Natural Sciences 
and Engineering Research Council of Canada~(NSERC), 
and TRIUMF, which receives federal funding via a contribution 
agreement with the National Research Council of Canada~(NRC). 
JDW is supported in part by the DOE under grant DE-SC0007859.

%%%%%%%%%%%%%%%%%%%%%%%%%%%%%%%%%%%%%%%%%%%%%%%%%%%%%%%%%%%%%%%%%%%%%%
%%%%%%%%%%%%%%%%%%%%%%%%%%%%%%%%%%%%%%%%%%%%%%%%%%%%%%%%%%%%%%%%%%%%%%

\bibliographystyle{JHEP}
\bibliography{gwstring_long}
%\bibliographystyle{JHEP}
%%%%%%%%%%%%%%%%%%%%%%%%%%%%%%%%%%%%%

\end{document}